\documentclass{tMPH2e}

\usepackage{amssymb}
\usepackage{amsmath}
\usepackage{mathabx}
\usepackage{graphicx}
\usepackage{amsbsy}
\usepackage{color}
\usepackage{bm}

\newcommand{\bq}{\begin{eqnarray}}
\newcommand{\eq}{\end{eqnarray}}
\newcommand{\bqn}{\begin{eqnarray*}}
\newcommand{\eqn}{\end{eqnarray*}}
\newcommand{\rr}{\mathbf{r}}
\newcommand{\dd}{\mathbf{d}}

\newcommand{\red}[1]{#1}

\begin{document}
\title{Wertheim perturbation theory: thermodynamics and structure of patchy colloids}

\author{Riccardo Fantoni$^{a}$  and Giorgio Pastore$^{b}$$^\ast$\thanks{$^\ast$Corresponding author. 
Email: 
pastore@ts.infn.it
\vspace{6pt}} \\\vspace{6pt} 
$^{a}${\em{ Dipartimento di Scienze Molecolari e Nanosistemi,
  Universit\`a Ca' Foscari Venezia, Calle Larga S. Marta DD2137,
  I-30123 Venezia, Italy
}};
$^{b}${\em{ Universit\`a di Trieste, Dipartimento di Fisica, strada Costiera 11, 34151
Grignano (Trieste), Italy
}}\\\vspace{6pt}\received{} }

\maketitle

\begin{abstract}
We critically discuss the application of the Wertheim's 
theory to classes of complex
associating fluids that can be today engineered in the laboratory as
patchy colloids and to the prediction of their peculiar gas-liquid
phase diagrams.\red{ Our systematic study,  stemming from 
perturbative version of the theory, allows us to show that, even at the simplest level of
approximation for 
the inter-cluster correlations, the 
theory is still able to provide a consistent and stable picture of the behavior of
interesting models of self-assembling colloidal suspension. We extend the analysis of 
a few cases of patchy systems recently introduced in the literature. In particular, 
we discuss  for the first time in detail the consistency of the structural   
description underlying the perturbative approach and 
we are able to prove a consistency relationship between the valence as
obtained from  thermodynamics and from the structure for the one-site case. A simple analytical
expression for the structure factor is proposed. }
\end{abstract}

\bigskip


\section{Introduction}
\label{sec:introduction}

Recently, there have been interesting developments of techniques for the
synthesis of new colloidal patchy particles in the laboratory
\cite{Yi2013}, including seeded growth, swelling, and phase
separation. Whereas in the laboratory relatively less work has been
done on the thermodynamic characterization of self-assembly 
of these particles, from a theoretical point
of view, or in recent computer experiments, these kind of associating
fluids \cite{Fantoni2014} and their clustering
and phase behavior are actively
studied \cite{Bianchi2006,Sciortino2007,Bianchi2008,Russo2011a,Russo2011b,Tavares2012,Rovigatti2013}. 

In principle, statistical mechanics should be able to describe all
equilibrium phases. However, the strong and confined attractions
responsible of association call for a more clever approach than brute
force. In particular, it has been found useful to describe an
associating fluid as one where there are $n_c$ 
species of clusters made of a number $i$ of particles, denoted 
$i$-mers. Many definitions of cluster are possible
\cite{Lee1973,Ebeling1980,Gillan1983,Caillol1995,Fisher1993,Friedman1979}
either of a geometric nature or of a topological one, depending on
the spatial arrangement of the bonded particles. If we measure the
concentrations of the $i$-mers in an 
associating fluid we will find that they are functions of the
thermodynamic state: For one-component systems, the temperature $T$ and the density $\rho$ of the
fluid.  Then, special statistical mechanics approaches have been
developed to obtain such information  and phase diagrams from models
of interactions.

In our previous work \cite{Fantoni2014} we compared two theories for
cluster equilibria, the Wertheim association
theory 
\cite{Wertheim1,Wertheim2,Wertheim3,Wertheim4} and the
Bjerrum-Tani-Henderson theory
\cite{Bjerrum1926,Tani83,Fantoni2011,Fantoni2012,Fantoni-Springer2013,Fantoni2013a,Fantoni2013b}
 and we showed that for $n_c=2$ the two approaches 
coincide when inter-cluster correlation are ignored, i.e. the system behaves as an ideal gas of clusters. 
Nonetheless, the
simple and elegant perturbation theory described in  Wertheim's work, is able, unlike the one of
Bjerrum-Tani-Henderson, to describe the case of  $n_c\to\infty$ fluids. Due to
this fact, Wertheim theory is able to describe the liquid phase, thus giving
access to the study of liquid-gas coexistence in a coherent way,
while the Bjerrum-Tani-Henderson one is not. 
The first 
order in the Wertheim perturbation theory approximation is a simple but very useful tool. 
At high temperature, the associating fluid reduces to
the ``reference'' fluid that can also be considered as the one
obtained from the associating fluid switching off all
attractions. However, in its original form, the theory is only
applicable when some ``steric incompatibility'' conditions are
fulfilled by the associating fluid: A single bond per site, no
more than one bond between any two particles, and no closed
loop, or ring, of bonds. 

Patchy colloids are systems of current experimental and theoretical \cite{Yi2013,Bianchi2011} 
interest. Simple models for their interactions, for example fluids of hard-spheres
decorated with attractive sites distributed on their surface, are well suited for 
application of Wertheim theory.
For particles with $M$ identical bonding sites, Bianchi {\sl et
al.} \cite{Bianchi2006,Sciortino2007,Bianchi2008} discovered the
``empty liquid'' scenario as $M$ approaches two, {\sl i.e.} when the
clusters allowed in the fluid are just the ``chains''. Even more rich 
phenomenology is found when there are sites of two different kinds 
\cite{Russo2011a,Russo2011b} and ``junctions'' formation becomes
possible. Such structures become
responsible for a re-entrance of the liquid branch of the binodal, and 
for ``rings'' formation
\cite{Tavares2012,Rovigatti2013}. Moreover, extending Wertheim
  theory beyond its steric incompatibility conditions, the rings
  formation has been found to be responsible for a re-entrance also in
  the gas branch and the appearance of a second lower critical point
  (recently appeared studies which further extend Wertheim theory to
  allow also for doubly bonded sites
  \cite{Marshall2012,Marshall2013,Marshall2013b}). From all these
  studies  
  emerged how Wertheim theory has very good semi-quantitative
  agreement with exact Monte Carlo simulations, when applied to these
  one-component patchy particle fluids (especially so at the level of
  the clusters concentrations behavior).
Far from being a purely theoretical speculation, these fluids can be
engineered in the laboratory \cite{Yi2013} from patchy colloids. 

In the present work, while critically reviewing such theoretical
results,  in particular elucidating the role of the accuracy of inter cluster correlations, 
we will discuss the solution of the Wertheim theory
applied to hard-spheres with $M$ identical bonding sites and 
with sites of two different kinds. 
Our analysis is intended to be as simple and systematic as
possible while re-analyzing the many works found in the literature on
various particular highly idealized associating colloidal suspension
models. This will allow us to treat the ring forming systems of Rovigatti 
{\sl et al.} \cite{Tavares2012,Rovigatti2013} fully analytically as
freely jointed chains. 
We show that also  the results in Ref. \cite{Tavares2012b}, extending 
 Russo {\sl et al.}
\cite{Russo2011a,Russo2011b} results to take into acccount the
``X-junctions'' formation, and in particular the existence of 
charcteristic  ``R'' shaped spinodals
are largely independent on the choice of the reference system 
correlations. Moreover, we find indication of
a gas-liquid coexistence with a critical point at extremely low
densities and temperatures at $r<1/3$, with $r$ the ratio between the
gain in energy between the bond of two unlike sites and the one
between two like sites.

We also study in detail the relationship between structural and thermodynamic information
within Wertheim theory, and in particular between the effective 
valence as obtained from the thermodynamics and from the structure.

The paper is organized as follows: In Section \ref{sec:thermodynamics}
we introduce the thermodynamic quantities we will take under
consideration in the rest of the work; in Section \ref{sec:theory} we
will review Wertheim association theory in the light of the present work 
needs, the problem of identical
attractive site (Section \ref{sec:MA}), and the problem of attractive
sites of two different kinds (Section \ref{sec:MA-MB}); in Section
\ref{sec:binodal} we introduce the problem of the gas-liquid
coexistence; in Section \ref{sec:microscopic} we comment on the
relevance of the pair-potential microscopic level of description; in
Section \ref{sec:results-w} we sistematically re-analyze many  results 
obtained applying Wertheim theory to specific fluids with identical
sites (Section \ref{sec:identical-sites}) and sites of two different
kinds (Section \ref{sec:different-sites}). We show, in a systematic way, that
all the results present in the literature are structurally 
stable with respect to changes in the 
reference system accuracy; 
in Section \ref{sec:rdf} we
determine a simple analytical expression for the radial distribution
function which we then use to calculate the valence; in Section
\ref{sec:sk} we determine a simple analytical expression for the
structure factor; Section \ref{sec:conclusions} is for final remarks. 

\section{Thermodynamics}
\label{sec:thermodynamics}

Consider a one-component fluid of $N$ associating hard-sphere (HS)
particles in a volume $V$ at an absolute temperature $T=1/\beta k_B$
with $k_B$ Boltzmann constant. 

The Helmholtz free energy $A$ of a hard-sphere associating fluid
can be written as a sum of separate contributions \cite{Jackson88} 
\bq \label{HS-mf-bond}
A=A_0+A_{mf}+A_{bond},
\eq 
where $A_\red{0}$ is the free energy of a  hard-sphere fluid at a density
$\rho=N/V$, $A_{mf}$ is the mean-field contribution due to the
dispersion forces, and $A_{bond}$ is the change in the free energy due
to association.
We will generally use the notation $a(\rho,T)=a=A/N$ for the free
energy per particle.

The hard-sphere free energy per particle in excess of the ideal gas
one is accurately given by the Carnahan and Starling expression
\cite{Carnahan69} 
\bq \label{Carnahan-Starling}
\beta a_0^{ex}=\frac{4\eta-3\eta^2}{(1-\eta)^2},
\eq
where $\eta=(\pi/6)\rho\sigma^3$ is the packing fraction of the
hard-spheres of diameter $\sigma$. So that adding the ideal gas
contribution $\beta a_{id}=\ln(\rho\Lambda^3/e)$, with $\Lambda$ the
de Broglie thermal wavelength, we obtain $a_0=a_{id}+a_0^{ex}$.

The mean-field contribution has the van der Waals form
\bq
\beta a_{mf}=-\frac{\epsilon_{mf}\rho}{k_BT},
\eq
where the constant $\epsilon_{mf}$ is the measure of the strength of
the mean-field attractions. The addition of this contribution to
$A_0$ is essential to have a gas-liquid coexistence.

From a microscopic point of view one can see, for example, the mean
field contribution as arising from the first order in $\beta$ in a high
temperature expansion of a thermodynamic perturbation theory treatment
of the square-well (SW) fluid, with the HS taken as the reference
system. So, the free energy of the corresponding associating fluid
will be given by $A=A_{SW}+A_{bond}$. But, as we will see in Section
\ref{sec:results-w}, one can have gas-liquid coexistence with just
$A=A_0+A_{bond}$ for a properly chosen $A_{bond}$.  

We can define a unit of length, ${\cal S}$, and a unit of
energy, ${\cal E}$, so that we can introduce a reduced density,
$\rho^*=\rho{\cal S}^3$, and a reduced temperature, 
$T^*=k_BT/{\cal E}$.

The association contribution $A_{bond}$ will be discussed in the next
section. 

\section{Associating fluids}
\label{sec:theory}

We recall here the main result of Wertheim association theory
\cite{Wertheim1,Wertheim2,Wertheim3,Wertheim4}. We write the bond
free energy per particle $a_{bond}$ such that the full 
free energy per particle of the associating fluid can be written as
$a=a_0+a_{bond}$, where $a_0$ is the contribution of the reference
fluid, the one obtained from the associating fluid setting to zero all
the bonding attractions. We discuss the importance of the choice of a proper
pair-potential for the fulfillment of the steric incompatibility
conditions in the microscopic description of the fluid. 
And we discuss
the problem of the determination of the gas-liquid coexistence line
(the binodal) in our one-component fluid.

\subsection{Wertheim statistical thermodynamic theory}
\label{sec:wertheim}

In  Wertheim theory \cite{Wertheim1,Wertheim2,Wertheim3,Wertheim4}
one assumes that each hard-sphere of the one-component fluid is
decorated with a set $\Gamma$ of $M$ attractive sites. 
Under the assumptions of: [i.] a single bond per site, [ii.] no
more than one bond between any two particles, and [iii.] no closed
loop, or ring, of bonds, one can write in a first order thermodynamic
perturbation theory framework, valid at reasonably high temperatures, 
\bq \label{bond-w}
\beta a_{bond}^{W}=\sum_{\alpha\in\Gamma}\left(\ln x_\alpha -
\frac{x_\alpha}{2}\right) + \frac{M}{2},
\eq 
where $x_\alpha=N_\alpha/N$ is the fraction of sites $\alpha$ that are
not bonded. We will also introduce the symbol $x_i$ to denote the
concentration of clusters made of a number $i$ of particles. We will
always use a Greek index to denote a specific site. We can solve for
the $x_\alpha$ from the ``law of mass action'' 
\bq \label{x-wertheim}
x_\alpha=\frac{1}{1+\rho\sum_{\beta\in\Gamma}x_\beta\Delta_{\alpha\beta}},
~~~\alpha\in\Gamma
\eq
where the probability to form a bond, once the available sites of the
two particles are chosen, is given by
$\rho\Delta_{\alpha\beta}=\rho\Delta_{\beta\alpha}$ and approximated as
\bq \label{Delta-w}
\Delta_{\alpha\beta}=\int g_0(r_{12})\langle
f_{\alpha\beta}(12)\rangle_{\Omega_1,\Omega_2} d\rr_{12}. 
\eq
Here $g_0$ is the radial distribution function of the
reference system, $f_{\alpha\beta}$ is the Mayer function between
site $\alpha$ on particle 1 and site $\beta$ on particle 2 (see
Section \ref{sec:microscopic}), and
$\langle \ldots \rangle_{\Omega_1,\Omega_2}$ denotes an angular
average over all orientations of particles 1 and 2 at a fixed relative
distance $r_{12}$. Eq. (\ref{x-wertheim}) should be solved for the
real physically relevant solution such that $\lim_{\rho\to 0}
x_\alpha=1$. \red{Even if we cannot exclude the possibility of having
  multiple solutions satisfying to this condition we never encountered
  such a case in the present work. Clearly we cannot assign any
  physical value to the branches with $x_\alpha
  \notin [0,1]$.}  

At high temperatures $\Delta_{\alpha\beta}\to 0$ and $x_\alpha\to 1$,
which means we have complete dissociation. At low temperatures
(Wertheim theory is a high temperature expansion but here we just
mean the formal low $T$ limit of the first order Wertheim results)
$\Delta_{\alpha\beta}\to\infty$ and $x_\alpha\to 0$, which means that
we have complete association. 

The number of attractive sites controls the physical behavior. Models
with one site allow only dimerization. The presence of two sites
permits the formation of chain and ring polymers. Additional sites
allow formation of branched polymers and amorphous systems.

\subsubsection{One attractive site}
\label{sec:M=1}

The case of a single attractive site was carefully considered in our
previous work \cite{Fantoni2014} where a comparison between the
Wertheim theory and the Bjerrum-Tani-Henderson theory
\cite{Bjerrum1926,Tani83,Fantoni2011,Fantoni2012,Fantoni-Springer2013,Fantoni2013a,Fantoni2013b}
was made.

\subsubsection{Identical attractive sites}
\label{sec:MA}

Another simple case we can consider in Wertheim theory is the one with
$M$ identical attractive sites of kind $A$ (we will always use a
capital letter to denote a site kind). Now the law of mass action
for $x=x_A$ (the fraction of unbonded specific sites of kind $A$) is
solved by 
\bq \label{xis}
x=\frac{2}{1+\sqrt{1+4M\rho\Delta}},
\eq
with $\Delta=\Delta_{AA}$. 

The free energy contribution due to association is now given by
\bq
\beta a_{bond}^{W}=M(\ln x-x/2)+M/2.
\eq

In this case $x_1=x^M$.

\subsubsection{Attractive sites of two kinds}
\label{sec:MA-MB}

A more complex case in Wertheim theory is the one with
$M_A$ identical attractive sites of kind $A$ and $M_B$ identical
attractive sites of kind $B$. Now the law of mass action
reduces to the following system of two coupled quadratic equations
\bq \label{xaxb1}
x_A+M_A\rho\Delta_{AA}x_A^2+M_B\rho\Delta_{AB}x_Ax_B&=&1,\\ \label{xaxb2}
x_B+M_B\rho\Delta_{BB}x_B^2+M_A\rho\Delta_{AB}x_Ax_B&=&1,
\eq
which admits in general a set of 4 different solutions for $(x_A,x_B)$
from which it is necessary to single out the physically relevant one.
In the event that there is no attraction between a site
of kind $A$ and a site of kind $B$ then $\Delta_{AB}=0$ and
the system simplifies to 
\bq \label{xca}
x_A&=&\frac{2}{1+\sqrt{1+4M_A\rho\Delta_{AA}}},\\ \label{xcb}
x_B&=&\frac{2}{1+\sqrt{1+4M_B\rho\Delta_{BB}}}.
\eq
In the event that there is no attraction between sites of the same
kind it simplifies to
\bq \nonumber
x_A&=&2/\{1+(M_B-M_A)\rho\Delta_{AB}+\\ \label{xda}
&&\sqrt{[1+(M_B-M_A)\rho\Delta_{AB}]^2+4M_A\rho\Delta_{AB}}\},
\eq
and $x_B$ obtained exchanging $A\leftrightarrow B$ in the equation
above. 

The free energy contribution due to association is now given by
\bq \nonumber
\beta a_{bond}^{W}&=&M_A(\ln x_A-x_A/2)+M_A/2+\\ \label{abondab}
&&M_B(\ln x_B-x_B/2)+M_B/2.
\eq

In this case $x_1=x_A^{M_A}x_B^{M_B}$.

\subsection{The gas-liquid coexistence}
\label{sec:binodal}

In order to determine the gas-liquid coexistence line (the binodal)
one needs to find the compressibility factor $z=\beta p/\rho$, with $p$
the pressure, and the chemical potential $\mu$ of the associating
fluid according to the thermodynamic relations
\bq \label{z}
z(\rho,T)&=&\rho\left(\frac{\partial\beta
  a}{\partial\rho}\right)_{T,N},\\ \label{bm}
\beta\mu(\rho,T)&=&\left(\frac{\partial\beta
  a\rho}{\partial\rho}\right)_{T,V}=z+\beta a.
\eq

The coexistence line is then given by the Gibbs equilibrium condition
of equality of the pressures and chemical potentials of the two phases
\bq \label{gibbs-1}
\rho_gz(\rho_g,T)&=&\rho_lz(\rho_l,T),\\ \label{gibbs-2}
\beta\mu(\rho_g,T)&=&\beta\mu(\rho_l,T),
\eq
from which one can find the coexistence density of the gas $\rho_g(T)$
and of the liquid $\rho_l(T)$ phases.

The critical point $(\rho_c,T_c)$ is determined by solving the
following system of equations
\bq \label{cp1}
\left.\frac{\partial z\rho}{\partial\rho}\right|_{\rho_c,T_c}=0,\\ \label{cp2}
\left.\frac{\partial^2 z\rho}{\partial\rho^2}\right|_{\rho_c,T_c}=0.
\eq

\subsubsection{The mean field case}
\label{sec:mf}

\begin{figure}[htbp]
\begin{center}
\includegraphics[width=8cm]{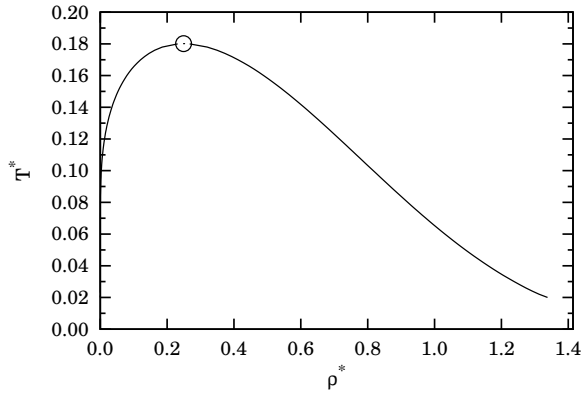}
\end{center}  
\caption{Gas-liquid binodal for the HS plus the van der Waals mean
  field term. The circle is the critical point at
  $\rho^*_c\approx 0.249129,T^*_c\approx 0.180155$, and $z_c\approx
  0.358956$ \cite{Hansen-McDonald-3}.} 
\label{fig:pd_mf}
\end{figure}

For the HS fluid in the presence of just a van der Waals mean field
free energy contribution, described by Eq. (\ref{HS-mf-bond}) 
without the last association term, the thermodynamics is parameter
free. We take the diameter of the spheres $\sigma$ as the unit of
length (so that $\rho^*\in [0,\sqrt{2}]$ with $\sqrt{2}$ the
close-packing reduced density) and $\epsilon_{mf}$ as the unit of
energy. Solving the Gibbs equilibrium conditions of
Eqs. (\ref{gibbs-1})-(\ref{gibbs-2}) we find the binodal of
Fig. \ref{fig:pd_mf} and from Eqs. (\ref{cp1})-(\ref{cp2}) we find the
critical point.   

We can see this case as describing a thermodynamic perturbation
theory approximation for a SW fluid to first order in $\beta$ small
\cite{Liu2009}. Monte Carlo simulations of the SW fluid are well known
to show a gas-liquid binodal with the critical point shifting at lower
temperatures and higher densities as the width of the attractive well
decreases \cite{Vega1992,Liu2005}. 

Recently \cite{Malescio2002} it was shown through numerical simulation
and theoretical approaches that a binodal with two maxima, implying
the existence of a low-density-liquid and a high-density-liquid, can
arise solely from an isotropic interaction potential with an
attractive part and with two characteristic short-range repulsive
distances. 

We consider the binodal of Fig. \ref{fig:pd_mf} as ``standard'' in the
sense that the gas branch $T_g(\rho)$ is a monotonously increasing
function of density and the liquid branch $T_l(\rho)$ a monotonously
decreasing function of density. We will see in the next section that
using Wertheim association theory it is possible to obtain non
standard binodals by replacing the mean-field contribution $A_{mf}$
with a proper association contribution $A_{bond}$. 
 
\subsection{Microscopic description: Importance of the pair-potential}
\label{sec:microscopic}

The fluid is assumed to be made of particles interacting only through
a pair-potential $\phi(12)=\phi(\rr_1,\Omega_1,\rr_2,\Omega_2)$ where
$\rr_i$ and $\Omega_i$ are the position vector of the center of
particle $i$ and the orientation of particle $i$ respectively.

To give structure to the fluid we further assume that the particles
have an isotropic hard-core of diameter $\sigma$ with
\bq \label{wertheim-potential}
\phi(12)=\phi_0(r_{12})+\Phi(12),
\eq
where $r_{12}=|\rr_{12}|=|\rr_2-\rr_1|$ is the separation between the
two particles 1 and 2 and  
\bq
\phi_0(r)=\left\{\begin{array}{ll}
+\infty & r\le\sigma\\
0       & r>\sigma
\end{array}
\right.,
\eq

The anisotropic part $\Phi(12)$ in Wertheim theory is generally chosen
as
\bq \label{site-site-potential}
\Phi(12)=\sum_{\alpha\in\Gamma}\sum_{\beta\in\Gamma}
\psi_{\alpha\beta}(r_{\alpha\beta}), 
\eq
where
\bq
\rr_{\alpha\beta}=\rr_2+\dd_\beta(\Omega_2)-\rr_1-\dd_\alpha(\Omega_1),
\eq
is the vector connecting site $\alpha$ on particle 1 with site $\beta$
on particle 2. Here $\dd_\alpha$ is the vector from the particle
center to site $\alpha$ with $d_\alpha<\sigma/2$. The site-site
interactions $\psi_{\alpha\beta}\le 0$ are assumed to be purely
attractive. The Mayer functions introduced in Section
\ref{sec:wertheim} are then  defined as
$f_{\alpha\beta}(12)=\exp[-\beta\psi_{\alpha\beta}(r_{\alpha\beta})]-1$.

Wertheim theory depends on the specific form of the site-site
potential only through the quantity $\Delta_{\alpha\beta}$ of
Eq. (\ref{Delta-w}), as long as the three conditions of a single bond
per site,  
no more than one bond between any two particles, and no closed loop of
bonds, are satisfied. A common choice, for example, is a square-well
form 
\bq \label{site-site-square-well}
\psi_{\alpha\beta}(r)=\left\{\begin{array}{ll}
-\epsilon_{\alpha\beta} & r\le d_{\alpha\beta}\\
0                    & r>   d_{\alpha\beta}
\end{array}
\right.,
\eq
where $\epsilon_{\alpha\beta}>0$ are site-site energy scales, the
wells depths, and $d_{\alpha\beta}$ are the wells widths. In this
case we must have 
$d_\alpha+d_\beta>\sigma-d_{\alpha\beta}$ moreover we will have 
\bq \label{Dab}
\Delta_{\alpha\beta}=K_{\alpha\beta}(\sigma,d_{\alpha\beta},\eta)
(e^{\beta\epsilon_{\alpha\beta}}-1).
\eq
We will also call $\lim_{\rho\to 0}K_{\alpha\beta}=K_{\alpha\beta}^0$
some purely geometric factors. Remember that $\lim_{\rho\to
0}g_0(r)=\Theta(r-\sigma)$ with $\Theta$ the Heaviside 
step function.

Another common choice is the Kern-Frenkel patch-patch pair-potential
model \cite{Kern03}. 

\section{Structural stability of Wertheim theory}
\label{sec:results-w}

There has recently been some relevant progress on the study of several
complex associating fluids through Monte Carlo (MC) simulations
and theoretically through the Wertheim theory outlined above. The comparison
between the two approaches shows
semi-quantitative agreement, between the exact MC results and the
approximated theoretical results, at the level of description of  clusters
concentrations and of gas-liquid binodal.
We will here return on some of the systems
studied from Bianchi {\sl et
al.} \cite{Bianchi2006,Sciortino2007,Bianchi2008}, Russo {\sl et
al.} \cite{Russo2011a,Russo2011b}, and Rovigatti {\sl et
al.} \cite{Tavares2012,Rovigatti2013} 
from a unified perspective, and 
concentrating ourselves on the
structural stability of the Wertheim theory, i.e. we will show that all the  
qualitative non standard features of the phase diagrams at a large extent do not depend on 
the accuracy of description of the reference system.

\subsection{Identical sites}
\label{sec:identical-sites}

The case of hard-spheres with a number $M$ of identical attractive
sites in various geometries on the surface of the spherical particle
has been studied by Bianchi {\sl et 
al.} \cite{Bianchi2006,Sciortino2007,Bianchi2008}. They showed that
the properties of the resulting fluid are largely independent from the
sites geometry \cite{Bianchi2008}. And the gas-liquid binodal  has a
liquid branch moving at 
lower densities as $M$ decreases. In particular the binodal vanishes
for $M\to 2$, a scenario that they called ``empty liquid'': The
critical temperature $T_c(M)$ and critical density $\rho_c(M)$ are
such that $\lim_{M\to 2}T_c=\bar{T}_c>0$ and $\lim_{M\to
  2}\rho_c=0$. There is then the formation of a homogeneous disordered
material at small densities below $\bar{T}_c$, {\sl i.e.} a stable equilibrium
gel. Moreover, in
their fluid with $M=2$, Bianchi {\sl et al.} observed linear
``chains'' formation: ``chaining''. 

This is quite different from what 
happens in fluids of \red{Kern and Frenkel patchy hard-spheres varying
the patches surface coverages \cite{Fantoni2007}. In
Ref. \cite{Fantoni2007} a study of criticality similar to the one of
Bianchi was made varying the attractive patch surface coverage
$\chi$.} As the surface coverage $\chi$ vanishes, $\lim_{\chi\to
  0}T_c=\lim_{\chi\to 0}\rho_c=0$ \red{was found in such cases}.

Liu {\sl et al.} \cite{Liu2009} repeated Bianchi study for a system of
square-wells (SW), \red{instead of HSs as in the Bianchi case}, with a
number $M$ of identical attractive sites. \red{In their study} the
gas-liquid coexistence remains also for $M\to 0$, \red{as expected in
  view of the comments of Section \ref{sec:mf}}.

\subsubsection{Gas-liquid binodal}
\label{sec:identical-sites-binodal}

With $M$ identical sites of kind $A$ we have in the site-site
interaction $\epsilon_{AA}=\epsilon$ which we take as unit of energy
and again we take $\sigma$ as unit of length. 

We now choose $a=a_0+a_{bond}$ with the association part given by
the Wertheim theory Eq. (\ref{bond-w}) with $M$ identical sites (see
Section \ref{sec:MA}).  

Following Ref. \cite{Sciortino2007} we choose the identical
sites distributed on the surface of the spherical particle and
\bq \label{dB}
d_{AA}=d= \left(\sqrt{5-2\sqrt{3}}-1\right)\sigma/2\approx
0.120\sigma,
\eq
which guarantees that each site is engaged at most in one
bond. Moreover we approximate the radial distribution function of the
reference system with its zero density limit taking
$\Delta_{AA}=\Delta=K^0\left[e^{\beta\epsilon}-1\right]$ and using, in  
Eq. (\ref{Dab}), the following expressions
\bq \label{average-f}
\langle f_{AA}(12)\rangle&=&
(e^{\beta\epsilon}-1)m_\red{AA}(r_{12})~~~r_{12}>\sigma,\\ \label{m(r)} 
m_\red{AA}(r)&=&\left\{\begin{array}{ll} \displaystyle
\frac{(d+\sigma-r)^2(2d-\sigma+r)}{6r\sigma^2} & \sigma<r<\sigma+d\\
0 & r>\sigma+d
\end{array}\right.,\\ \nonumber
K^0_{AA}=K^0&=&4\pi\int_\sigma^{\sigma+d}m_{AA}(r)r^2dr\\ \label{K0B}
&=&\pi d^4(15\sigma+4d)/30\sigma^2\\ \nonumber
&\approx& 0.332\times 10^{-3}\sigma^3.
\eq

\begin{figure}[htbp]
\begin{center}
\includegraphics[width=8cm]{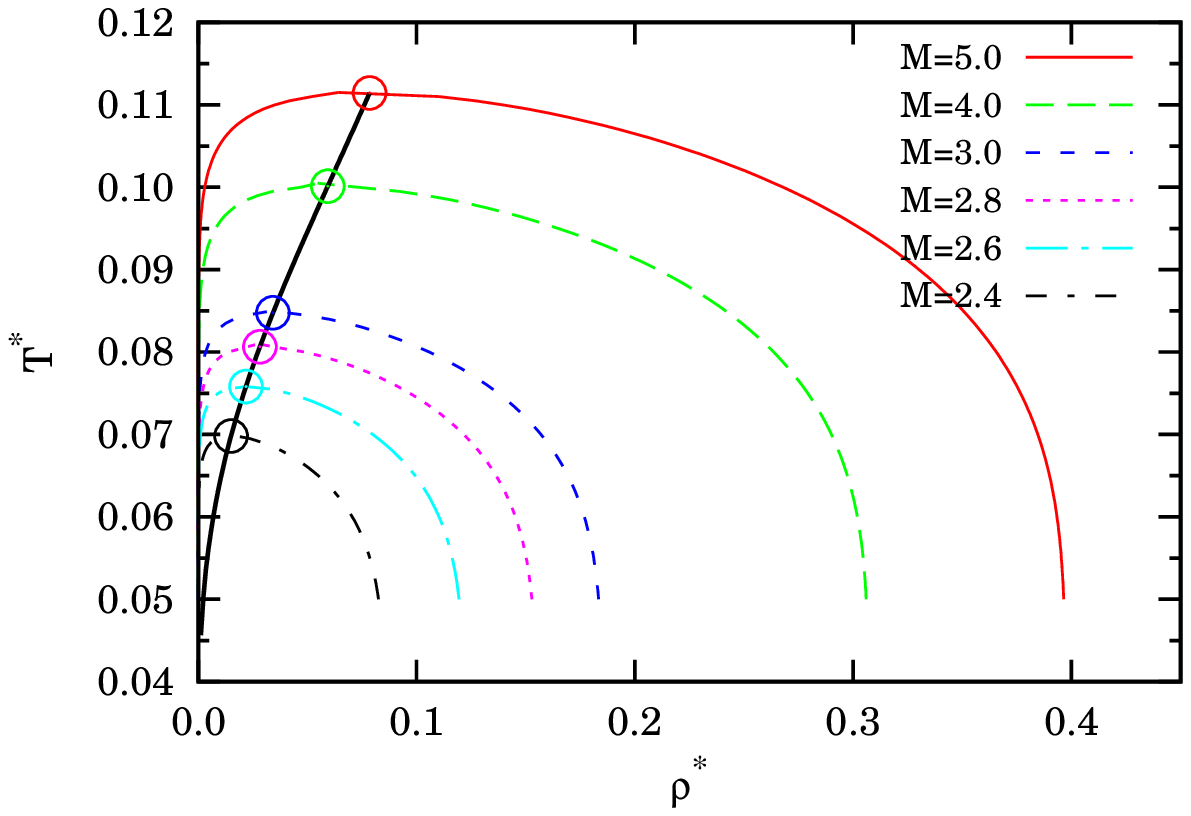}\\
\includegraphics[width=8cm]{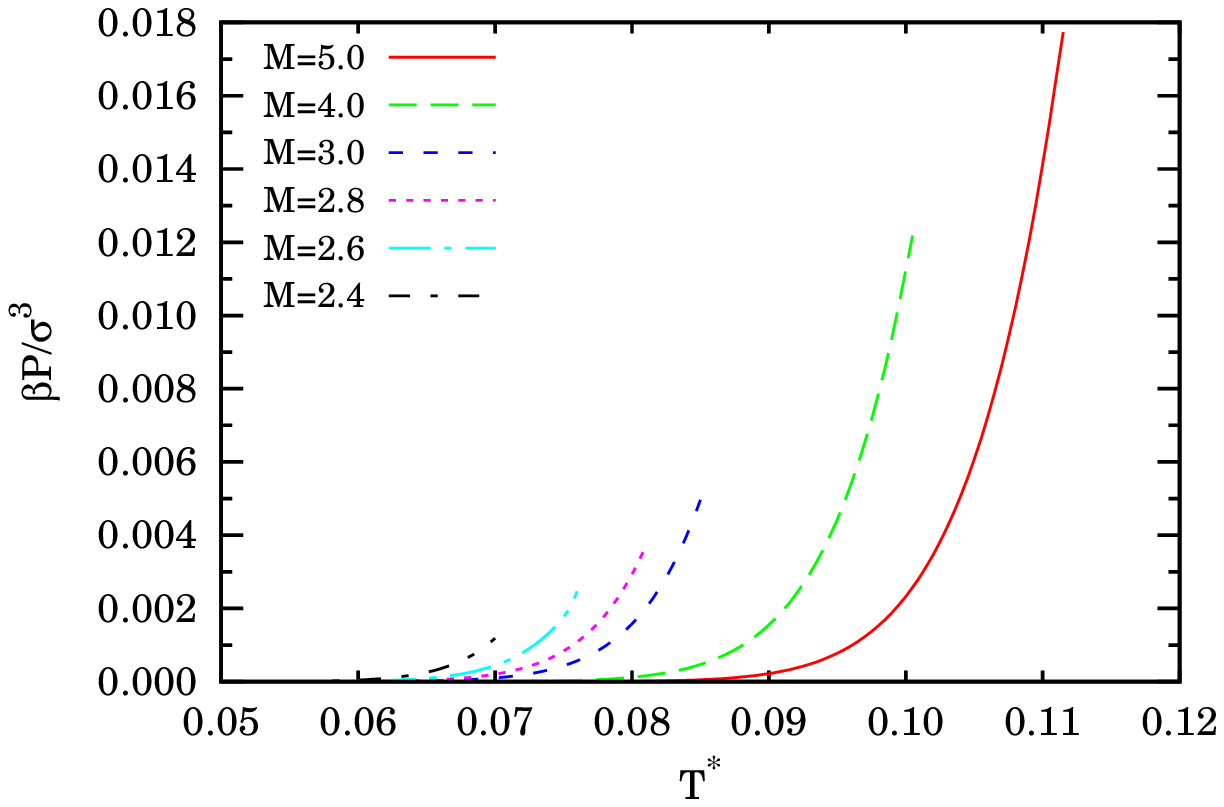}
\end{center}  
\caption{(color online) Top panel: Evolution of the gas-liquid binodal
  as a function of $M$. The continuous thick black line is the locus
  of the critical points for $M\in ]2,5]$. Bottom panel:
  Pressure-temperature diagram.} 
\label{fig:pd_m}
\end{figure}

In Fig. \ref{fig:pd_m} we show the evolution of the gas-liquid binodal
as a function of $M$, the only free parameter in Wertheim thermodynamic
perturbation theory. Compared with Fig. 4 of Bianchi {\sl et al.}
\cite{Bianchi2006} we see how the qualitative behavior stays 
the same even if the two figures differ slightly quantitatively due
to our further approximation of taking the radial distribution of
the reference system equal to one in the range where bonding
occurs. This shows how the Wertheim theory is robust in its
qualitative phase diagram predictions. The binodal appears to be
always a standard one. And, as we can see from the figure, upon
approaching $M\to 2$ the coexistence disappears.  Bianchi {\sl et al.}
\cite{Bianchi2006} called this 
phenomenon the empty liquid scenario. It in particular tells us that the
fluid with $M=2$, with the two sites chosen at the spherical particle
poles in order to avoid the formations of rings (closed loops of
bonds), is made only by chains and does not admit a gas-liquid
coexistence. The non-integer $M$ cases can be realized through a
binary mixture \cite{Bianchi2006,Heras2011,Heras2011b}. 

From the point of view of Wertheim theory the reason for this scenario
can be explained simply by looking at the low temperature limit for
the bond contribution to the pressure
\bq \nonumber
\beta p_{bond}^W&=&\rho z_{bond}^{W}=\rho^2\frac{\partial\beta
  a_{bond}^{W}}{\partial\rho}\\
&=&
-\frac{2M^2\Delta\rho^2}{\left(1+\sqrt{1+4M\Delta\rho}\right)^2}
\stackrel{\Delta\to\infty}{\longrightarrow}-\frac{M}{2}\rho.
\eq
From which immediately follows that for $M>2$ the pressure as a
function of density on a low temperature isotherm shows a van
der Waals loop at low densities, which implies the occurrence of a
gas-liquid coexistence region.

\subsection{Sites of two kinds}
\label{sec:different-sites}

Tavares {\sl et al.} \cite{Tavares2009,Tavares2010} studied the case
of HS with three sites, two identical $A$ sites at the poles and a
third $B$ one. In addition to chaining, here they observe the
formation of ``junctions'': ``branching''; 
\red{rings formation is inhibited in these cases since the $A$ sites
  at the poles have very small well widths and the $B$ site position
  is chosen so as to avoid small bond loops, {\sl i.e.} triangular and
  square arrangements of bonded particles}. 
Two types of junctions are possible 
in models where $AA$ bonds are responsible for the chaining:
X-shaped junctions, due to $BB$ bonds, and Y-shaped junctions,
due to $AB$ bonds. They found that when two of the three interaction
strengths vanish simultaneously, there can be no liquid-vapor
coexistence. These correspond to the limits of non interacting linear 
chains ($\epsilon_{AA}\neq 0, \epsilon_{BB}=\epsilon_{AB}=0$), dimers
($\epsilon_{BB}\neq 0, \epsilon_{AA}=\epsilon_{AB}=0$), and hyperbranched
polymers ($\epsilon_{AB}\neq 0,\epsilon_{AA}=\epsilon_{BB}=0$) of
Eq. (\ref{xda}). \red{They also showed that the phase transition
always disappears as $\epsilon_{AA}\to 0$}. 
Moreover they showed that whereas ``X-junctions'' only yield a
critical point if their formation is energetically favorable, fluids
with ``Y-junctions'' will exhibit a critical point, even if forming
them raises the energy, provided this increase is below a certain
threshold. 

Russo {\sl et al.} \cite{Russo2011a,Russo2011b} extended Tavares study to
the case of two identical small $A$ sites at the poles and nine
equispaced identical big $B$ sites on the equator. Killing the
interaction between two $B$ sites ($\epsilon_{BB}=0$) they observed
the formation of chains and Y-junctions (and possibly hyperbranched
polymers for $\epsilon_{AB}/\epsilon_{AA}$ large enough) and
eventually a re-entrant behavior of the liquid branch of the
gas-liquid binodal pinched at low temperatures. 

Rovigatti {\sl et al.} \cite{Tavares2012,Rovigatti2013} extended Russo
model selecting an off-pole position of the $A$ sites, thus adding the
possibility of ``rings'' formation, and observed re-entrance both
in the gas and in the liquid branch of the binodal with a second lower
critical point where the coexistence curves closes itself at low
temperatures without the pinch. They needed to relax assumption
[iii.] in Wertheim theory \cite{Sear1994,Galindo2002,Avlund2011}.

\subsubsection{Gas-liquid binodal}
\label{sec:different-sites-binodal}

Russo {\sl et al.} \cite{Russo2011a} studied the case of sites of two
different kinds when the site-site interaction is restricted to 
$\epsilon_{BB}=0$ (no X-junctions). Then choosing as unit of
energy $\epsilon_{AA}$ and again $\sigma$ as the unit of length the
Wertheim theory depends on only five parameters:
$r=\epsilon_{AB}/\epsilon_{AA}>0$ and $M_A, M_B, K_{AA}, K_{AB}$. 

We now choose $a=a_0+a_{bond}$ with the association part given by
the Wertheim theory Eq. (\ref{bond-w}) with sites of two different
kinds (see Section \ref{sec:MA-MB}). In particular with the condition
$\epsilon_{BB}=0$, Eqs. (\ref{xaxb1})-(\ref{xaxb2}) admit just a set
of 3 different solutions for $(x_A,x_B)$ from which it is necessary to
single out the real physically relevant one such that $\lim_{\rho\to
  0}x_A=\lim_{\rho\to 0}x_B=1$. 

\begin{figure}[htbp]
\begin{center}
\includegraphics[width=8cm]{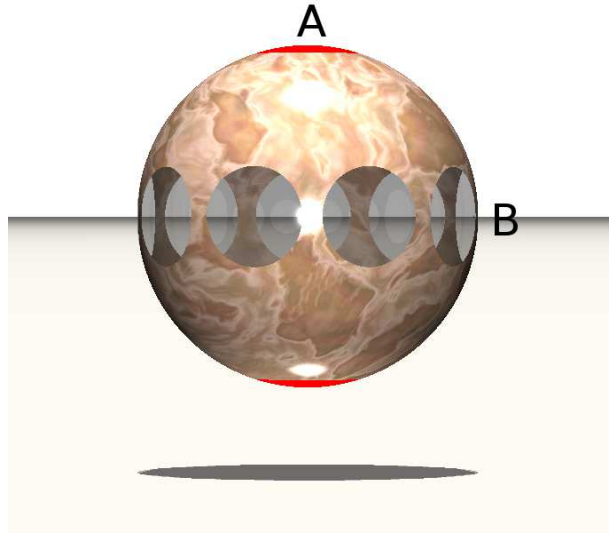}
\end{center}  
\caption{(color online) Pictorial view of a colloidal particle with
  attractive sites of two different kinds: Two $A$ sites on the poles
  and nine $B$ sites on the equator.}
\label{fig:colloid}
\end{figure}

Following Ref. \cite{Russo2011a} we choose $M_A=2, M_B=9$ (see
Fig. \ref{fig:colloid}) and 
$K^0_{AA}=1.80\times 10^{-4}\sigma^3, K^0_{AB}=1.56\times
10^{-2}\sigma^3$. In order to fulfill the Wertheim condition [i.], of a
single bond per site, the small $A$ sites are meant to reside at the
particle poles and the big $B$ sites equispaced on the particle
equator. The choice of $K^0_{AA}\ll K^0_{AB}$ and the large $M_B$ make
branching entropically favorable. We then approximate
$\Delta_{AA}=K^0_{AA}(e^{\beta \epsilon_{AA}}-1)$ and
$\Delta_{AB}=K^0_{AB}(e^{\beta \epsilon_{AB}}-1)$.  

\begin{figure}[htbp]
\begin{center}
\includegraphics[width=8cm]{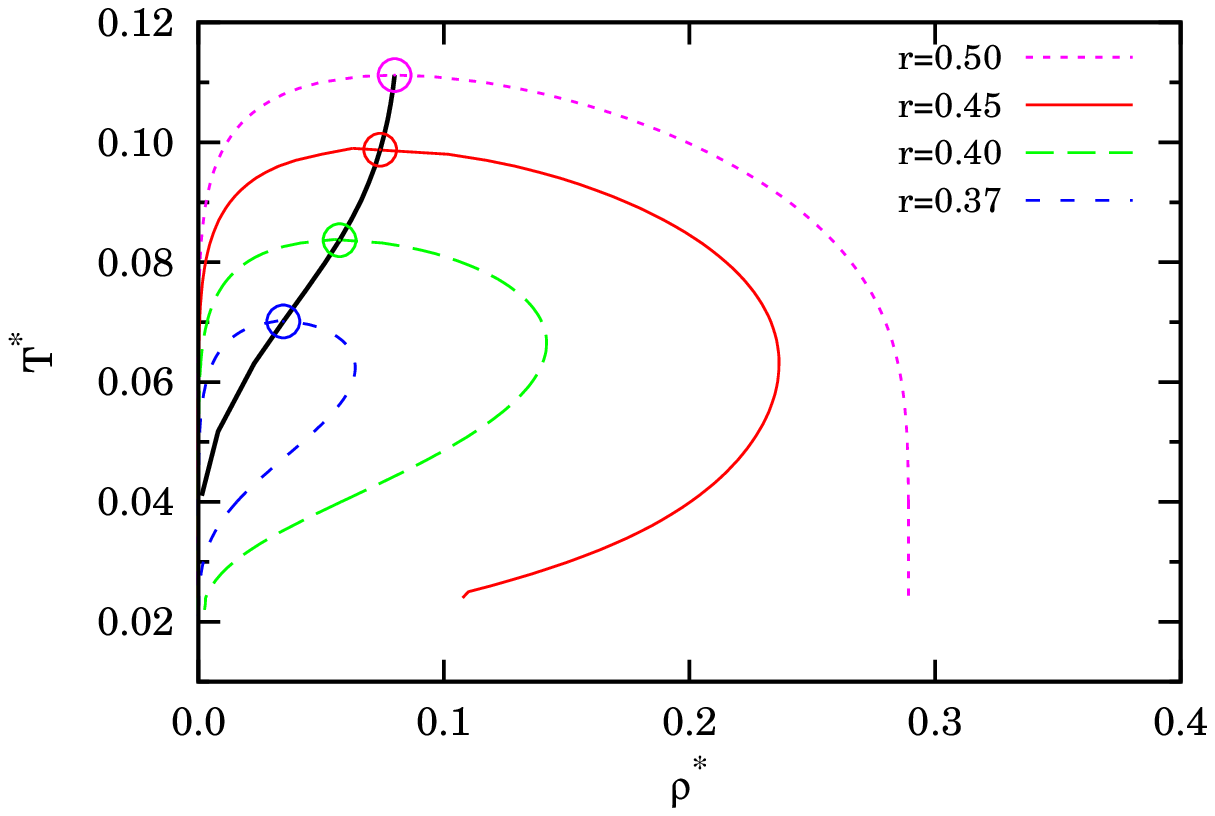}\\
\includegraphics[width=8cm]{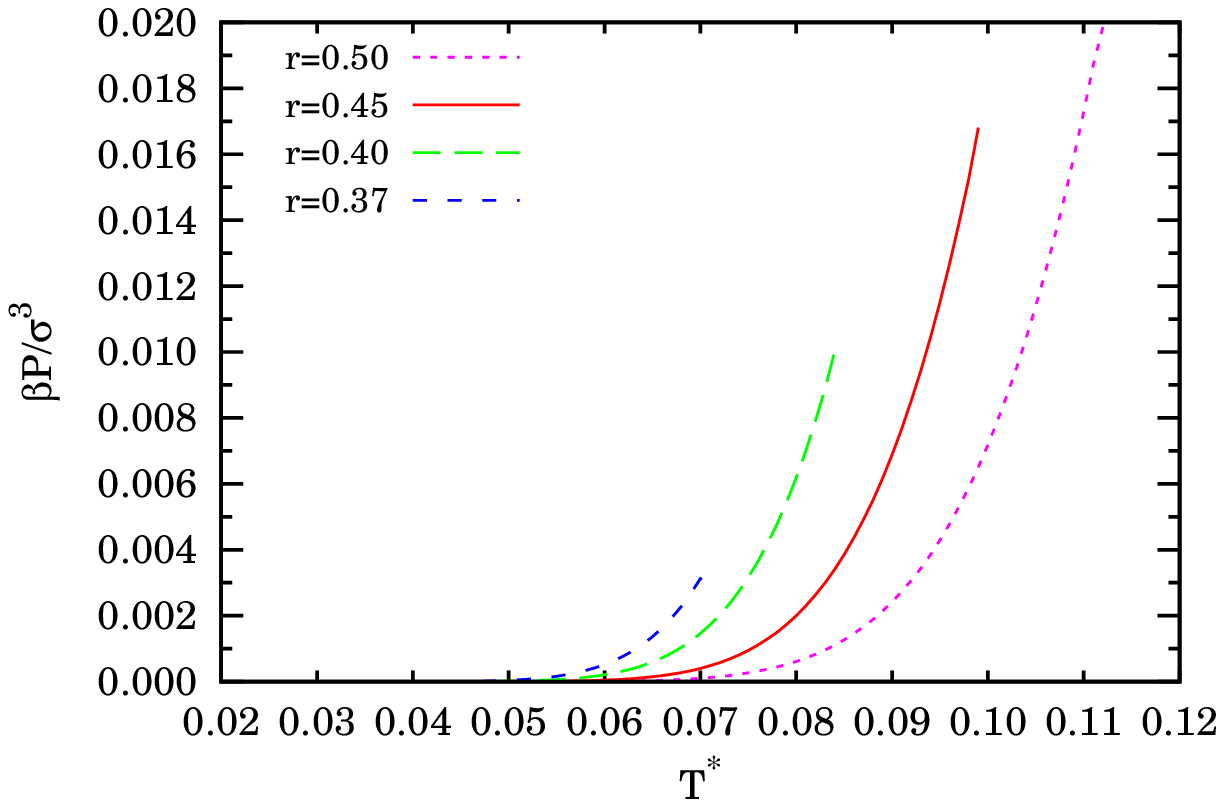}\\
\includegraphics[width=8cm]{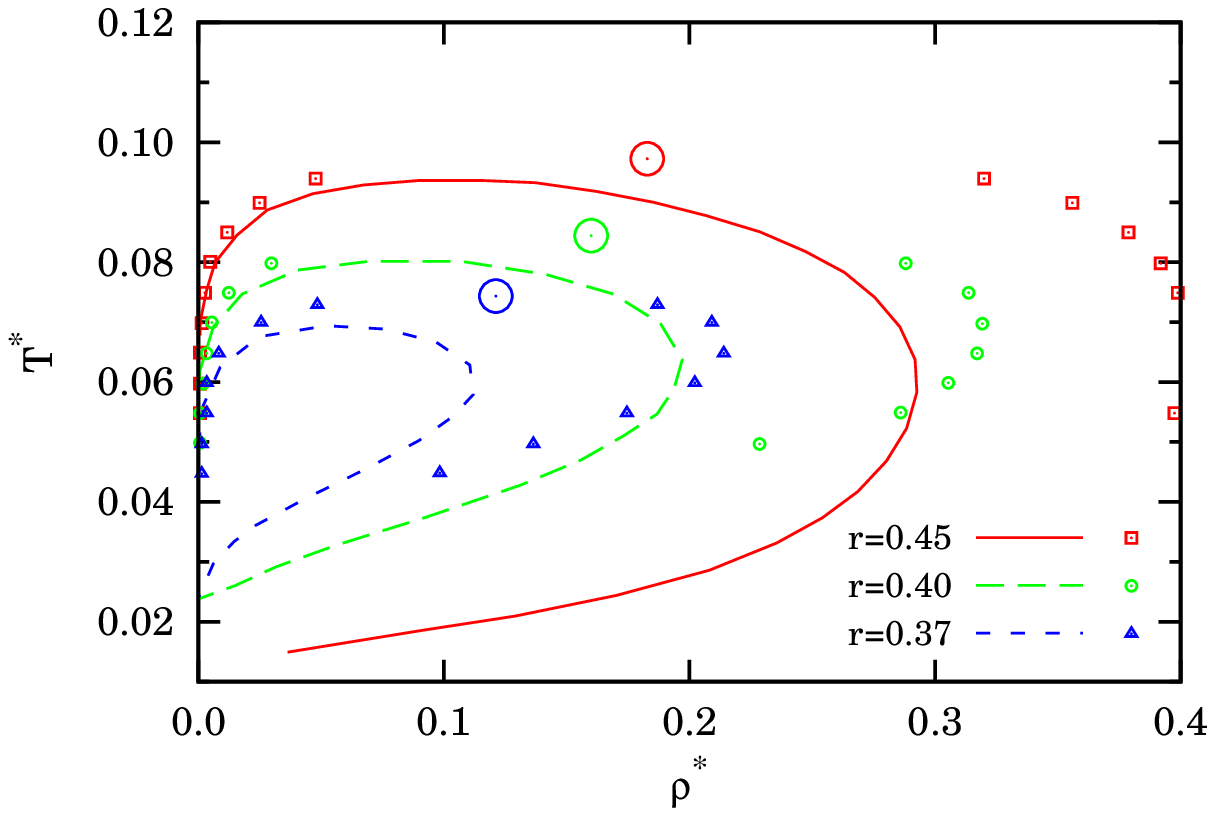}
\end{center}  
\caption{(color online) Top panel: Evolution of the gas-liquid binodal as a
  function of $r$. The continuous thick black line is the locus of the
  critical points for $r\in ]1/3,1/2]$. Middle panel:
  Pressure-temperature diagram. Bottom Panel: binodals of Russo {et al.}
  \cite{Russo2011b} Fig. 4 as obtained from their analysis (lines) of
  the Wertheim theory and from their MC simulations (points); the big
  circles are their predicted critical points.}  
\label{fig:pd_f}
\end{figure}

In Fig. \ref{fig:pd_f} we show the evolution of the gas-liquid binodal
as a function of $r$. Once again, comparing with Fig. 3 of Russo
{\sl et al.} \cite{Russo2011a} we observe a complete qualitative
agreement, even if in our calculation we further approximated the
radial distribution of the reference system equal to one independently
of density. We see that for $r<1/2$ we have a non standard 
binodal with a re-entrant liquid branch and a ``pinched'' shape
evidence that indeed the topological phase separation of Tlusty and
Safran \cite{Tlusty2000} is observed. Russo {\sl et al.} \cite{Russo2011a} 
were able to provide a qualitative  explanation for  this behavior by
analyzing the energetic of the junction formation process: since the
energy cost of forming a chain end is
$\epsilon_{chain}=\epsilon_{AA}/2>0$ and the energy cost of forming a
Y-junction is 
$\epsilon_{Y-junction}=-\epsilon_{AB}+\epsilon_{AA}/2=\epsilon_{AA}(1/2-r)$,
for $r<1/2$ we have $\epsilon_{Y-junction}<0$, and at low temperatures
only chains, which we already saw that do not phase separate, are present.  

They are also able to conclude that  phase separation
occurs only if $r > 1/3$. For $r < 1/3$, the energy cost
of forming junctions being too high or, alternatively, the entropy
gain being too small to offset the loss of translational entropy of
chains in the liquid phase.

\begin{figure}[htbp]
\begin{center}
\includegraphics[width=6cm]{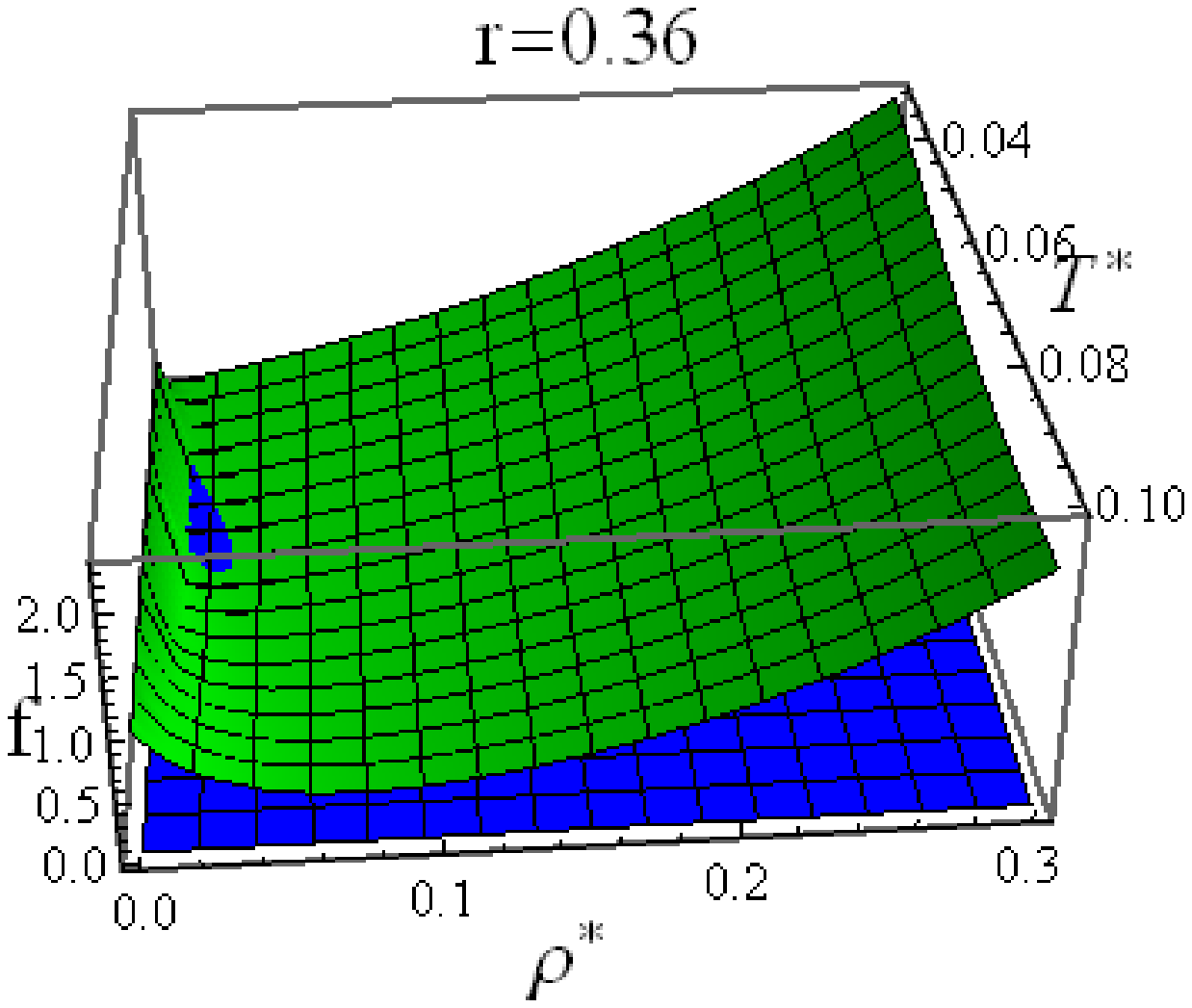}\\
\includegraphics[width=6cm]{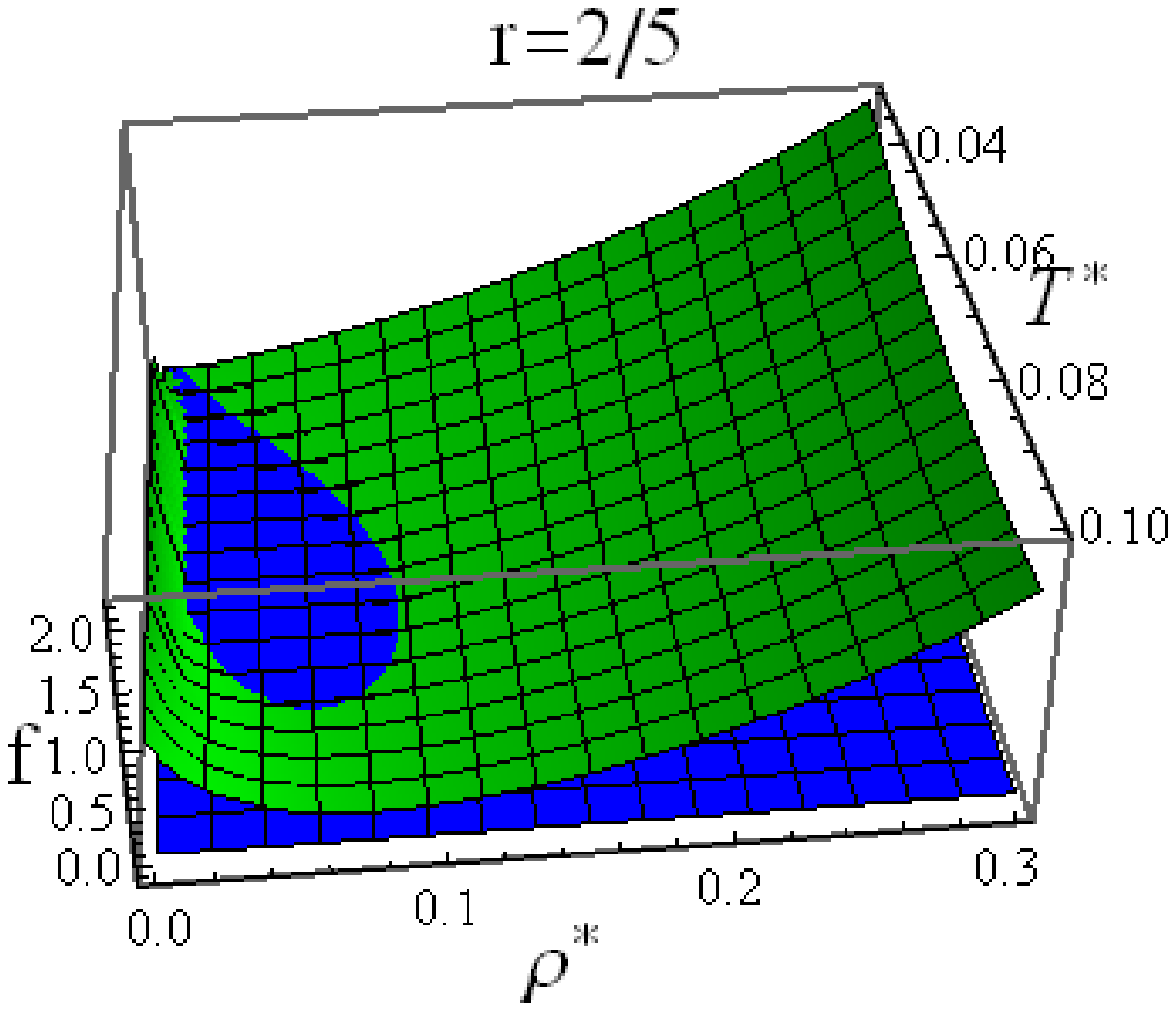}\\
\includegraphics[width=6cm]{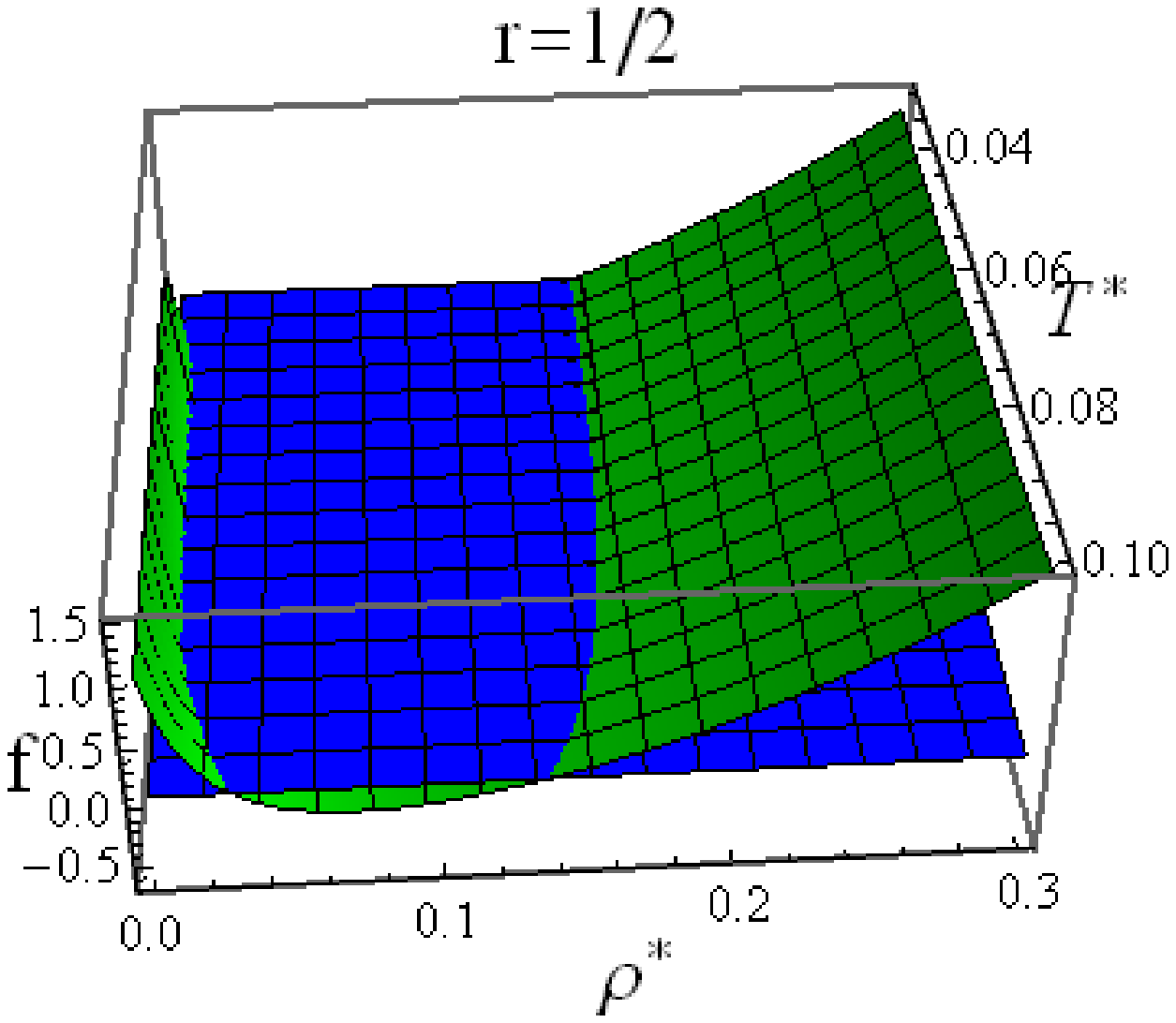}
\end{center}  
\caption{(color online) Tridimensional plots of $f(T,\rho;r)=d\beta
p/d\rho$ (green surface) for $r=0.36,2/5,1/2$ from top to bottom. Also
shown is the plane $f=0$ (blue surface). For $r=1/3$ the two surfaces
become tangent at small temperatures and small densities. For $r>
1/2$ the minimum in the pressure moves at larger densities at smaller
temperatures.}
\label{fig:3d}
\end{figure}

This behavior can be understood by looking at $f(T,\rho;r)=d\beta
p/d\rho=d\beta(p_0+p_{bond}^W)/d\rho$. Differently from Bianchi {\sl et
al.} case now we have $\lim_{\rho\to 0}d\beta p_{bond}^W/d\rho=0$. The
zeroes of $f$ are two lines in the $(\rho,T)$ plane, one for the
minima of the pressure and one for the maxima. The union of the two
lines is called the spinodal line for the coexistence. The equal area 
construction tells us that the binodal line encloses the spinodal line
and the two lines are tangent at the critical point. In
Fig. \ref{fig:3d} we show a tridimensional plot of $f$ for
$r=0.36,2/5,1/2$ as a function of temperature and density. Clearly the
three different scenarios do not depend on the specific values of
$K_{AA},K_{AB},M_A,M_B$ which only influence the region in the phase
diagram $(\rho,T)$ where we have the van der Waals loop.  

The cluster populations for the chain ends, $2x_A$, and Y-junctions, 
$9(1-x_B)$, along the binodal were studied in Ref. \cite{Russo2011a}
and are shown in their Fig. 4. From Fig. 9 of Ref. \cite{Russo2011b}
we see how the mean value of the number of bonds per particle (the
valence), $2(1-x_A)+9(1-x_B)$, tends to 2 at low temperatures, {\sl
  i.e.} the fluid tends to be formed essentially by chains which, in
agreement with Bianchi {\sl et al.} analysis, are unable to sustain
the gas-liquid coexistence.

The study of Russo {\sl et al.} differs substantially from the Janus
fluid case
\cite{Sciortino2009,Fantoni2011,Fantoni2012,Fantoni-Springer2013} 
where it is found a re-entrant gas branch for the gas-liquid binodal.

Rovigatti {\sl et al.} \cite{Rovigatti2013} extended Russo study to take
account of rings formation. In this case the expression for the
Wertheim bond free energy per particle of Eq. (\ref{abondab}) with
$M_A=2$ should be corrected as follows
\bq \nonumber
\beta a_{bond}^{W}&=&\ln\left(yx_B^{M_B}\right)-x_A-\frac{M_B}{2}x_B+\\
&&1+\frac{M_B}{2}-\frac{G_0}{\rho},
\eq
where $G_n$ is the $n$th moment of the rings size distribution
\bq \label{Gn}
G_n=\sum_{i=i_{min}}^\infty i^n W_i (2\rho\Delta_{AA}y)^i,
\eq
here $i_{min}$ is the minimum ring size, $y$ is the fraction of 
particles with the two $A$ sites unbonded,
and $W_i$ is the number of configurations of a ring of size
$i$. Assuming for the rings the freely jointed chain level of
description we can approximate \cite{Sear1994}
\bq \label{treloar}
(i+1)W_{i+1}=\frac{i(i-1)}{8\pi}\sum_{j=0}^l\frac{(-1)^j}{j!(i-j)!}
\left(\frac{i-1-2j}{2}\right)^{i-2},
\eq
for $l$ the smallest integer which satisfies $l\ge
(i-1)/2-1$. Expression (\ref{treloar}) is due to Treloar \cite{Flory}
and is the value of the end-to-end distribution function for a freely
jointed chain of $i$ links, when the end links are the length of one
link apart (the link length is equal to the diameter of a sphere which
we take to be our unit of length). For $i\gg 1$ it has the following
asymptotic behavior \cite{Flory}
\bq
(i+1)W_{i+1}\approx\left(\frac{3}{2\pi
  i}\right)^{3/2}e^{-3/2i},~~~i\gg 1,
\eq

The laws of mass action of Eqs. (\ref{xaxb1})-(\ref{xaxb2}), for
$\epsilon_{BB}=0$,  should now be corrected to take into account of
the $G_n\neq 0$ as follows 
\bq \label{rma1}
x_A^2&=&y(1-G_1/\rho),\\ \label{rma2}
1-x_A&=&M_B\rho\Delta_{AB}x_Bx_A+2\rho\Delta_{AA}x_A^2+G_1/\rho,\\ \label{rma3}
1-x_B&=&2\rho\Delta_{AB}x_Ax_B.
\eq
Note that solving for $x_A$ Eq. (\ref{rma1}) and for $x_B$
Eq. (\ref{rma3}) and substituting into Eq. (\ref{rma2}) one finds an
equation in $y$ only, which always admits just one solution $\bar{y}$
with the properties $0\le\bar{y}\le 1$ and $\lim_{T\to 0}\bar{y}=0$.

\begin{figure}[htbp]
\begin{center}
\includegraphics[width=8cm]{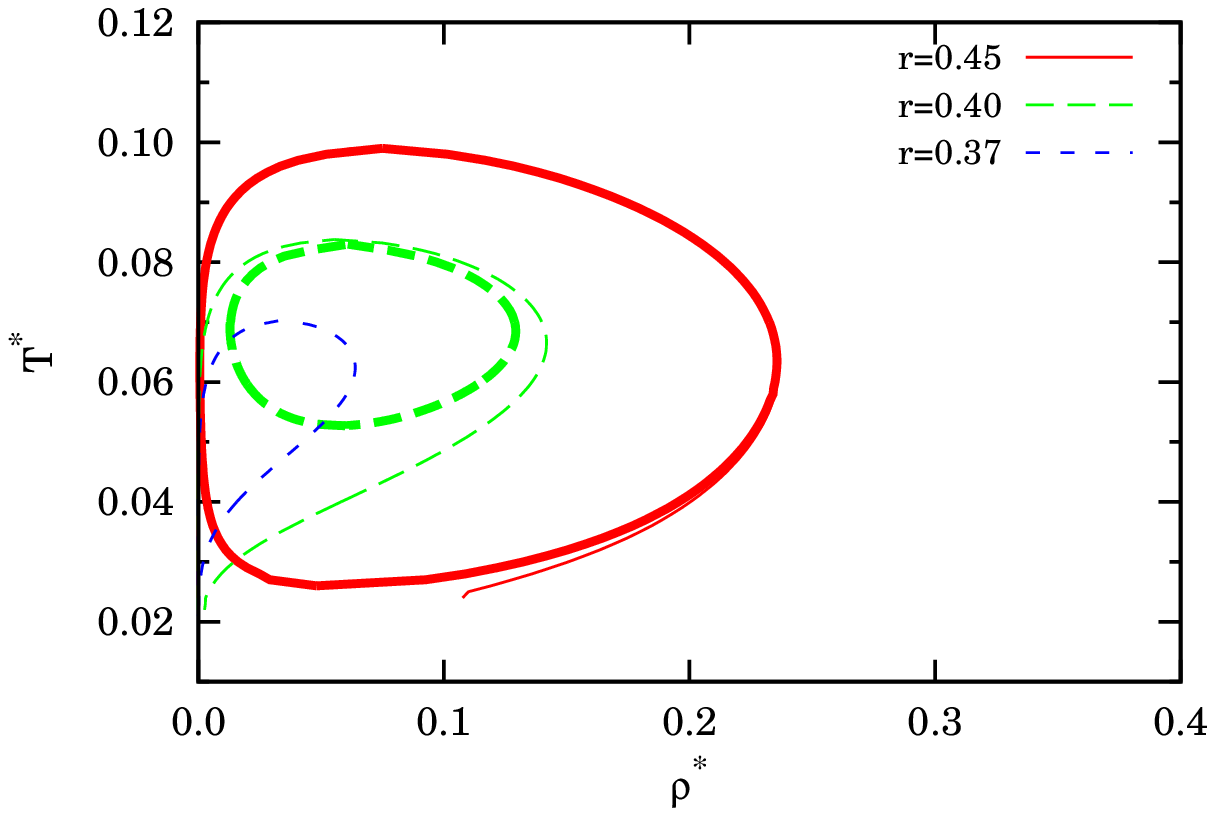}\\
\includegraphics[width=8cm]{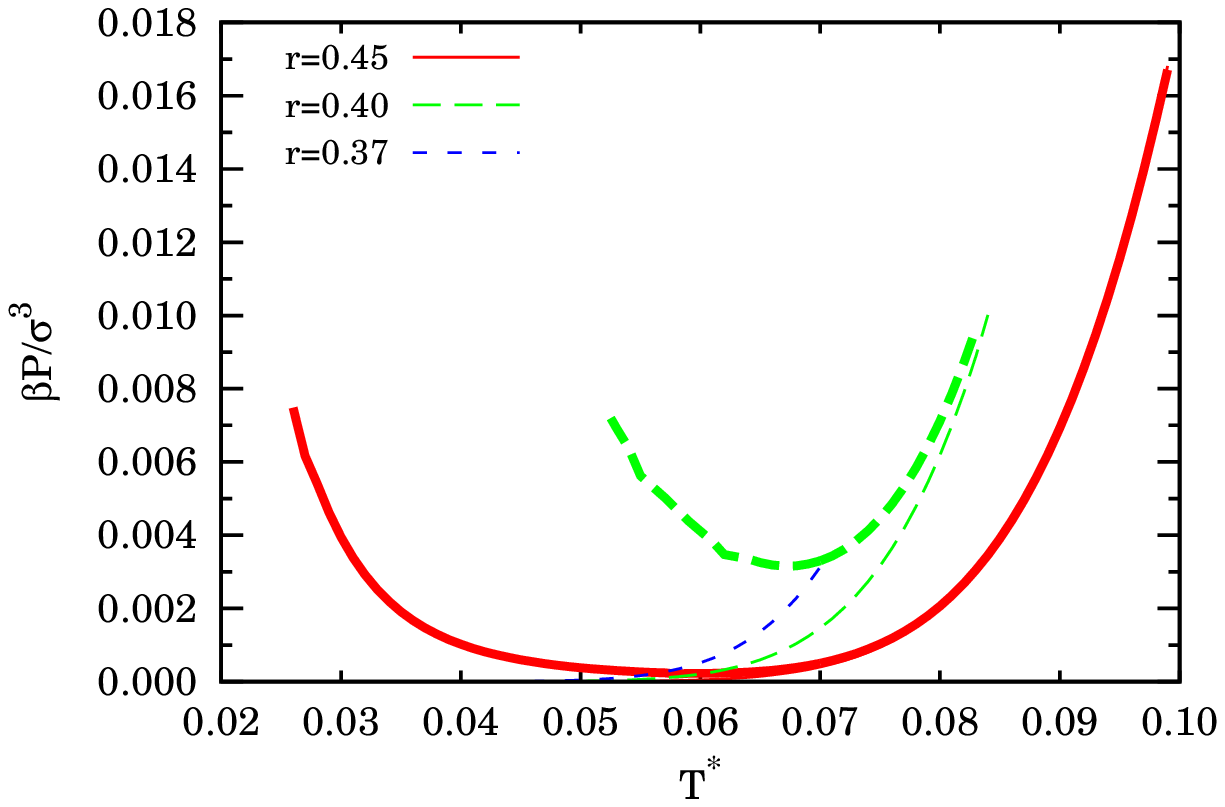}
\end{center}  
\caption{(color online) Top panel: Evolution of the gas-liquid binodal
  as a function of $r$. The thin lines are the binodals of
  Fig. \ref{fig:pd_f}. The thick lines are the results obtained for
  the rings forming fluid. Bottom panel: Pressure-temperature diagram.} 
\label{fig:pd_r}
\end{figure}

In Fig. \ref{fig:pd_r} we show our theoretical numerical results for
the gas-liquid binodal of the ring forming fluid. A comparison
with Fig. 1 of Rovigatti {\sl et al.} \cite{Rovigatti2013} shows
again a good qualitative agreement between the two calculations. In
our calculation we retained the first 50 terms in the convergent
series of Eq. (\ref{Gn}) and chose
$M_B=9$ and $\Delta_{AA},\Delta_{AB}$ as before. As we can
see the rings formation is responsible for the re-entrance in both the
gas and liquid branches of the binodal and for the appearance of a
second lower critical point. At $r=0.37$ we could not find a
coexistence line, leaving a system for which self-assembly is the
only mechanism for aggregation. 

In particular upon approaching the upper critical point, at $T=T_c^u$,
if we make a reversible transformation going from the liquid phase to
the vapor phase on an isotherm, at $T<T_c^u$, we will have, as usual
\bq
\Delta S=\int \frac{\delta Q}{T}=\frac{\lambda_vm}{T}>0,
\eq 
with $\Delta S$ the change in entropy $S=-(\partial A/\partial T)_{N,V}$,
$\delta Q$ the infinitesimal heat exchanges along the path of the
transformation, $\lambda_v$ the ``latent'' heat 
of vaporization, and $m$ the mass of the fluid.
Whereas Rovigatti {\sl et al.} \cite{Rovigatti2013} show that upon
approaching the lower critical point, at $T=T_c^l$, in the same
transformation at $T>T_c^l$, one finds 
\bq
\int\frac{\delta Q}{T}=\frac{\lambda_vm}{T}=\Delta S<0,
\eq
so that the ``latent'' heat of vaporization changes sign as $T$ varies
from $T_c^u$ to $T_c^l$. This can be seen directly from our
pressure-temperature diagram of Fig. \ref{fig:pd_r} using the
Clapeyron-Clausius formula \cite{FermiT}.
 
Rovigatti analysis neglects the rings with $AB$ bonds. We think that
their inclusions may have dramatic effects on the phase diagram.

\subsubsection{A possible extension}
\label{sec:extension}

\begin{figure}[htbp]
\begin{center}
\includegraphics[width=6cm]{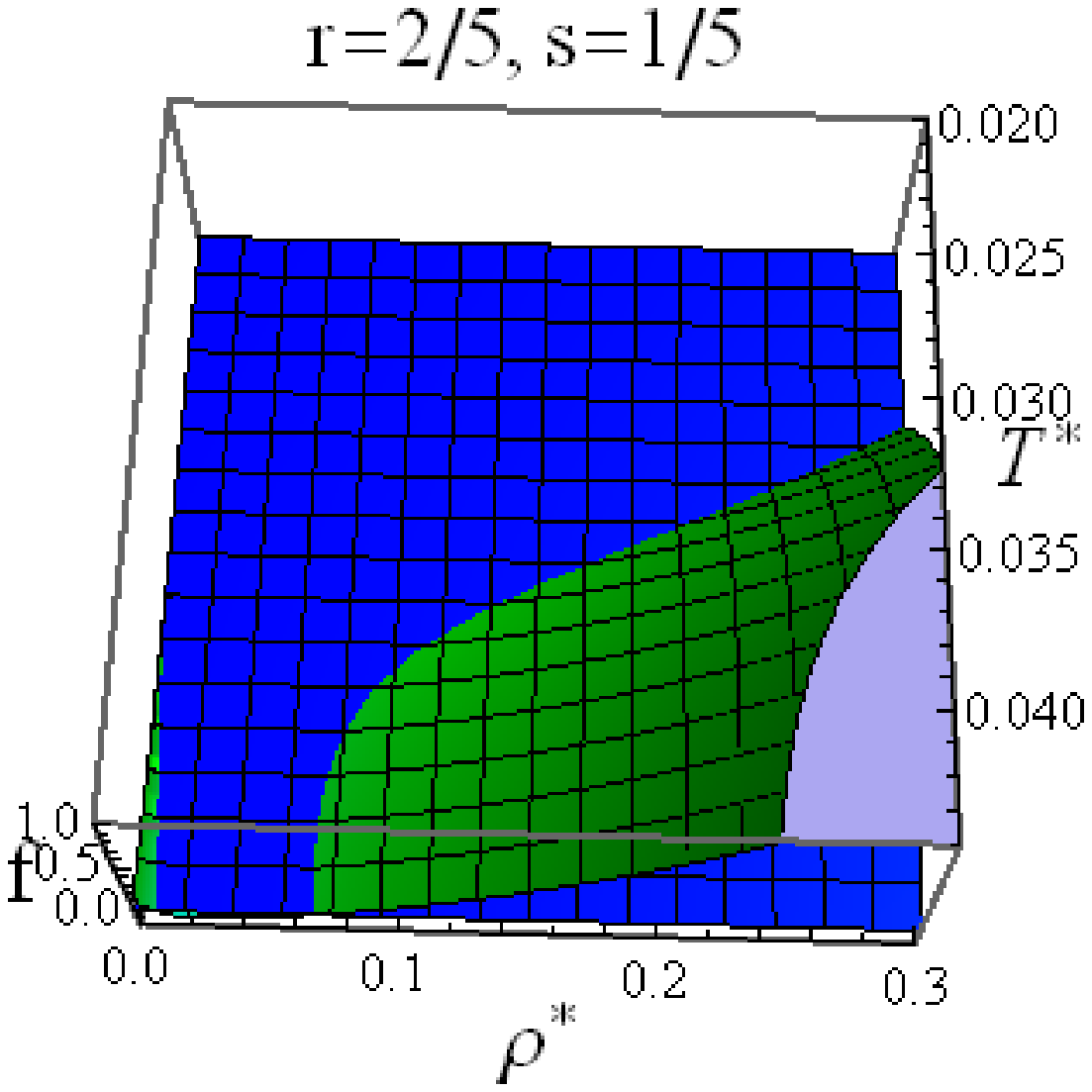}\\
\includegraphics[width=6cm]{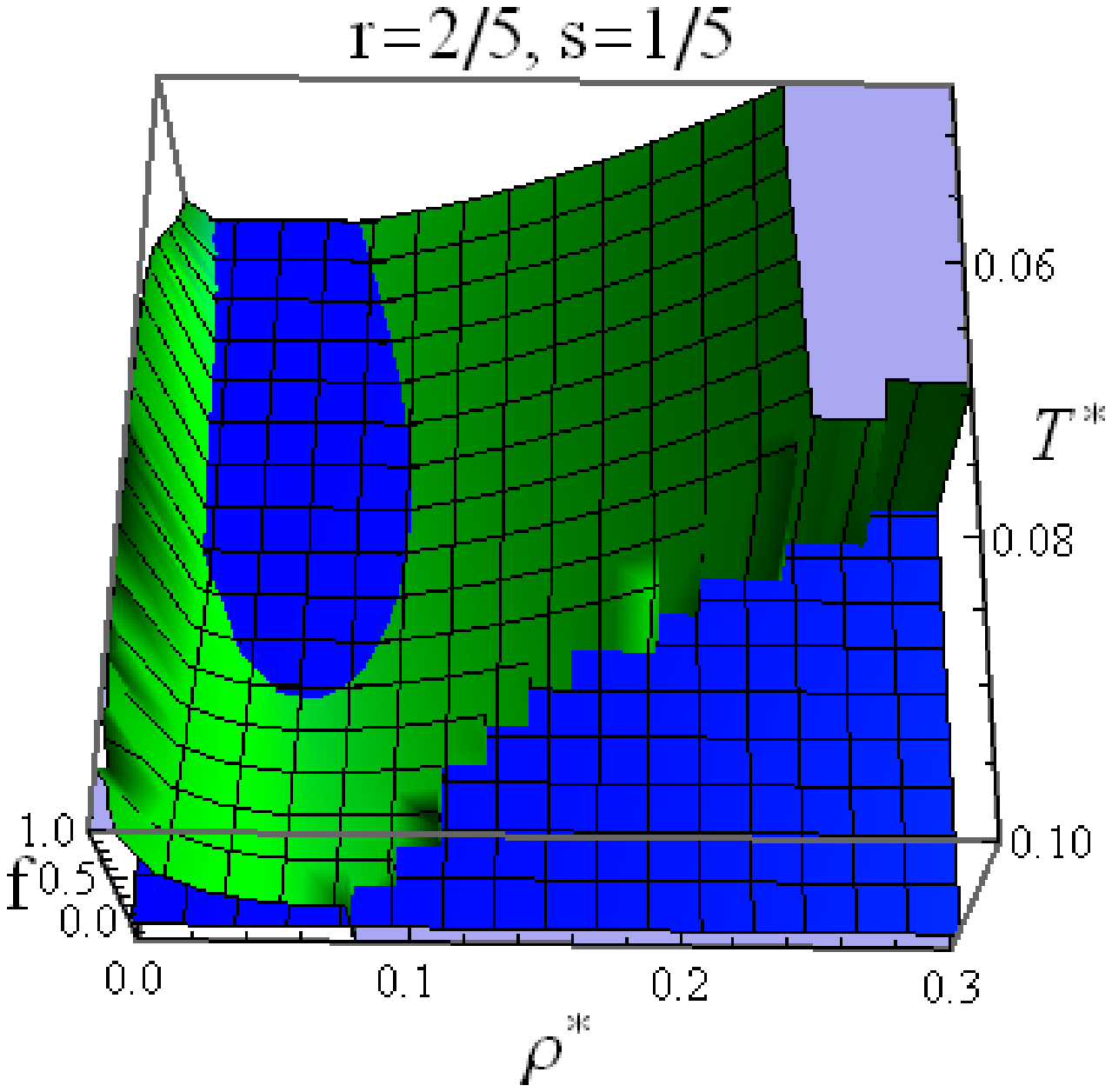}
\end{center}  
\caption{(color online) Tridimensional plots of
    $f(T,\rho;r,s)=d\beta p/d\rho$ (green surface) and of the plane
    $f=0$ (blue surface) for $(r,s)=(2/5,1/5)$. We show two plots one
    at high temperature and one at low temperature because the
    $(x_A,x_B)$ physical solution determination changes in the two
    regions of the phase diagram. The negative $f$ in the high
    temperature and high density corner of the lowest plot is due to
    another change in the physical solution determination.} 
\label{fig:rs1}
\end{figure}

\begin{figure}[htbp]
\begin{center}
\includegraphics[width=6cm]{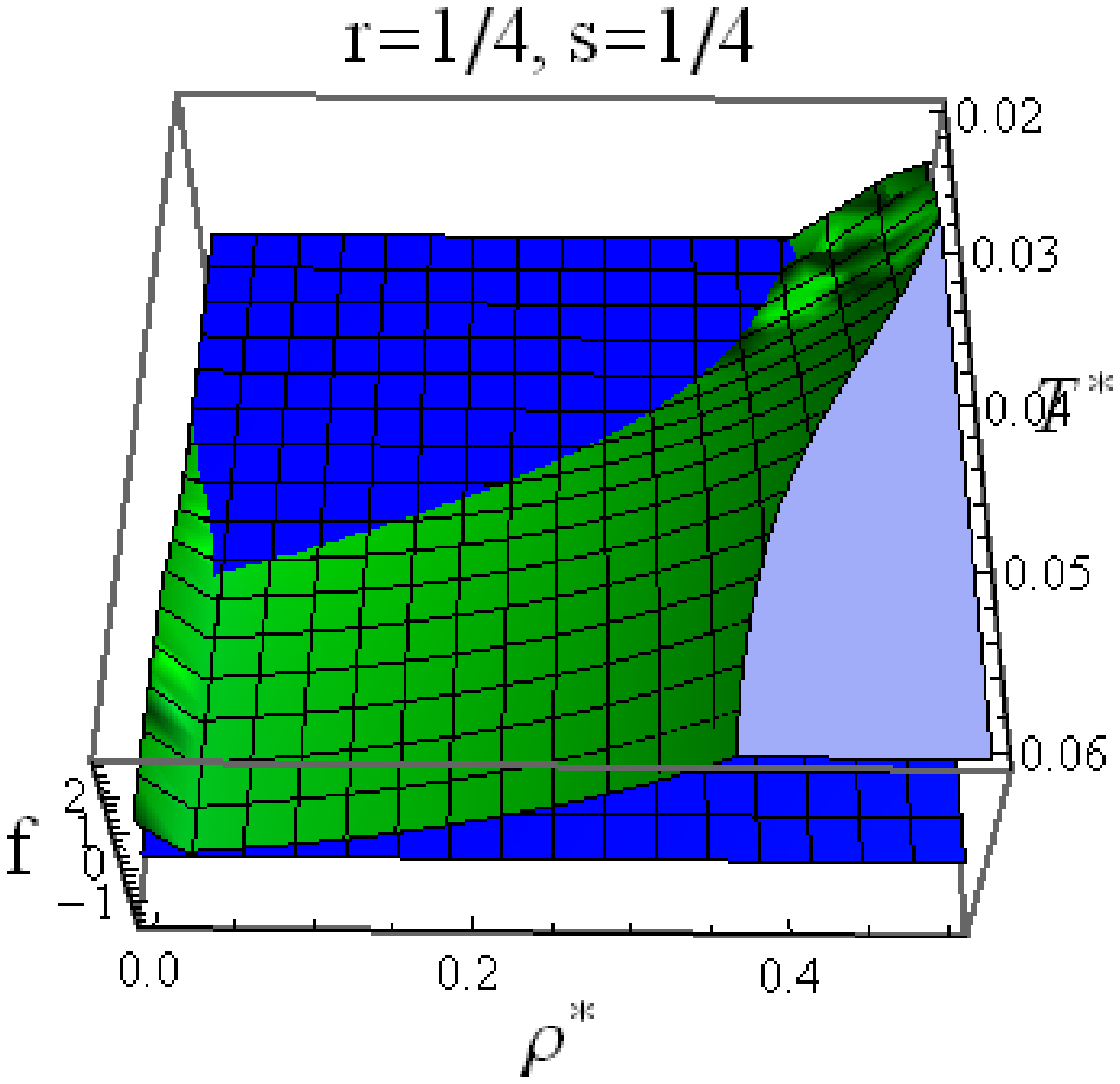}\\
\includegraphics[width=6cm]{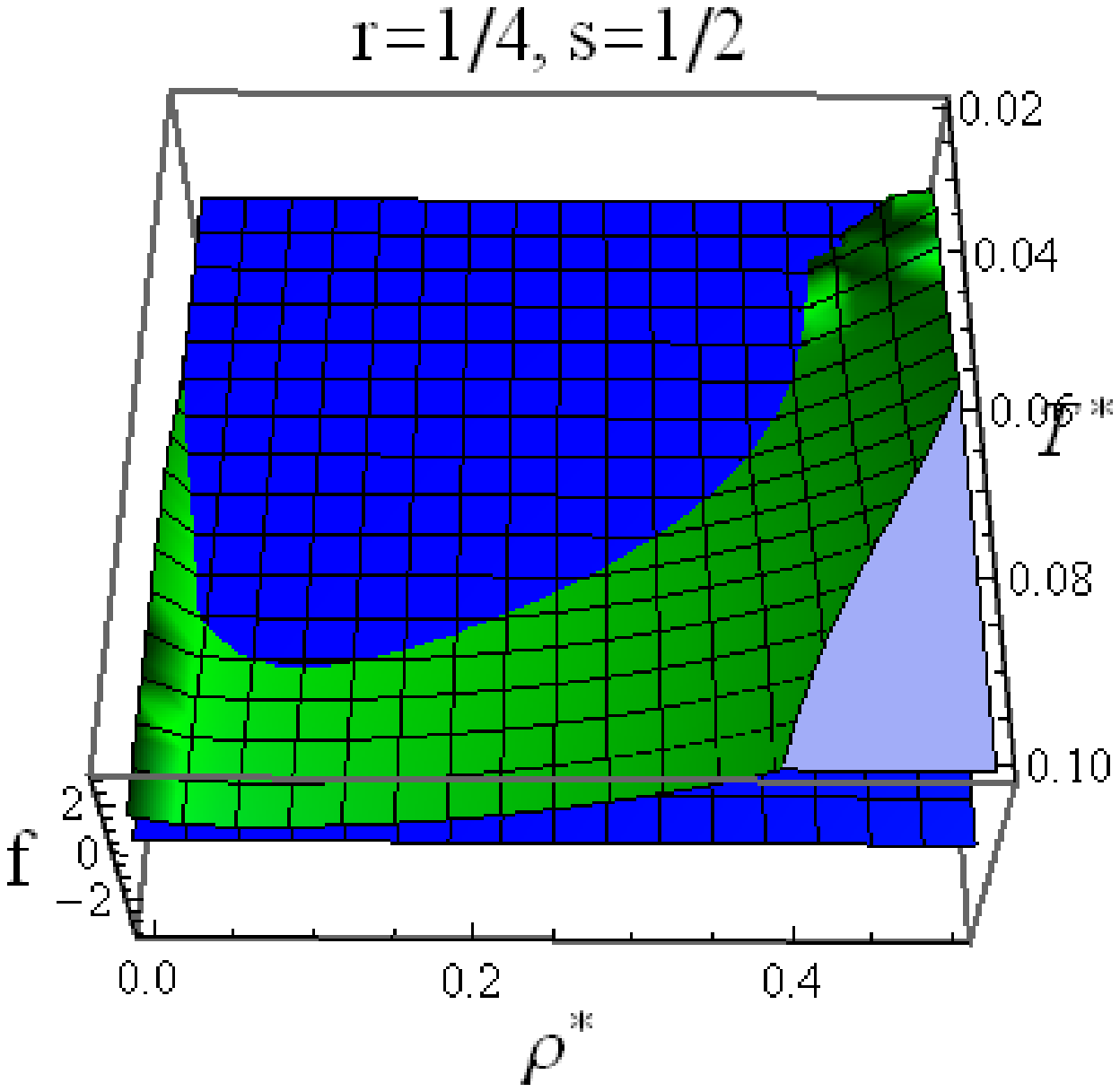}\\
\includegraphics[width=6cm]{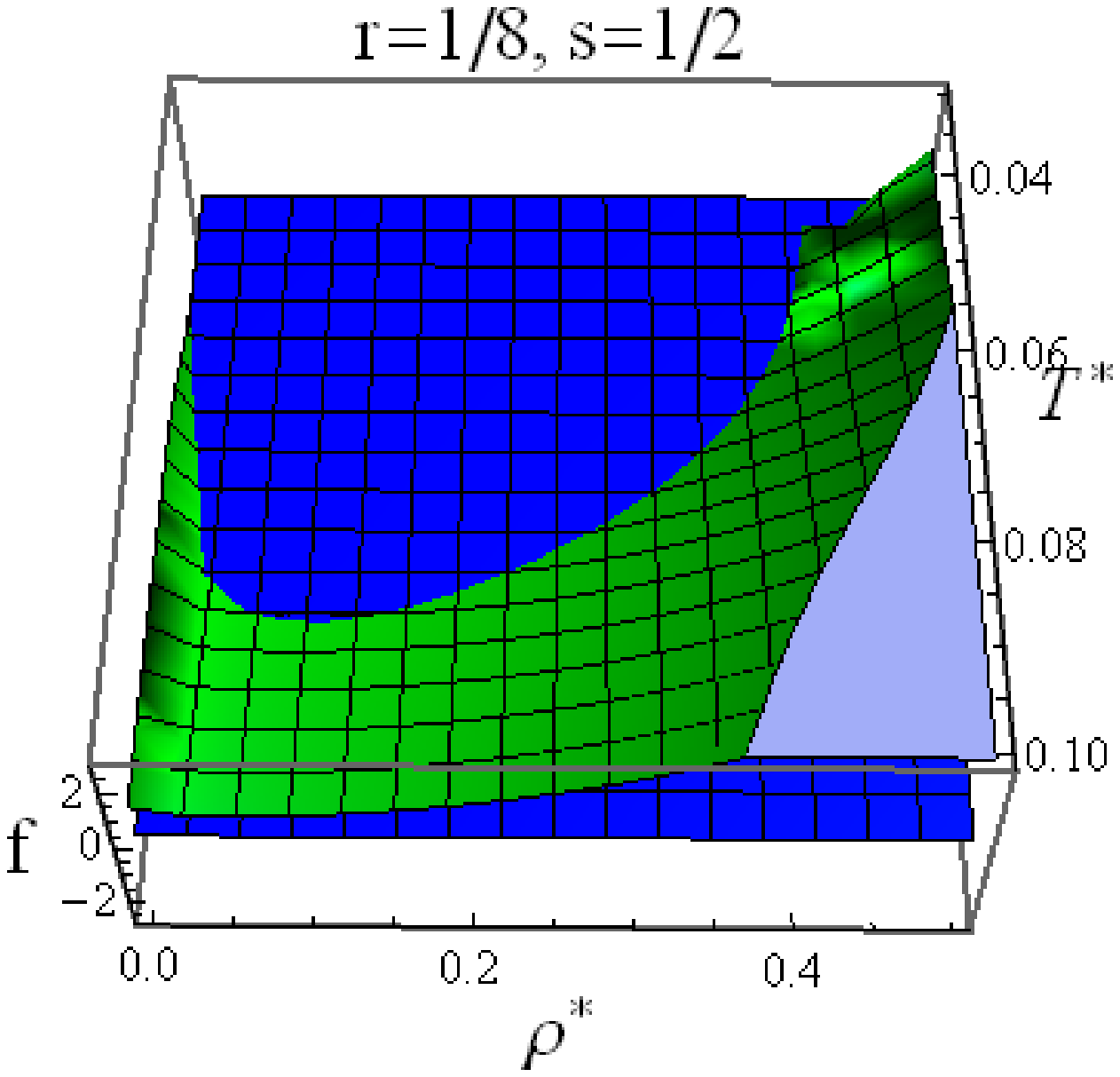}
\end{center}  
\caption{(color online) Tridimensional plots of
$f(T,\rho;r,s)=d\beta p/d\rho$ (green surface) for
$(r,s)=(1/4,1/4), (1/4,1/2), (1/8,1/2)$. 
Also shown is the plane $f=0$ (blue surface). As we can
see the spinodals of the two cases $(r,s)=(1/4,1/2),(1/8,1/2)$ look
essentially the same.}
\label{fig:rs2}
\end{figure}

It is possible to extend Russo {\sl et al.} \cite{Russo2011a,Russo2011b}
results allowing for the $\epsilon_{BB}\neq 0$ condition, responsible
for the X-junctions formation \cite{Tavares2012b}. \red{The analysis
  for just three sites, two of kind $A$ and one of kind $B$, can be
  found in Refs. \cite{Tavares2009,Tavares2010} were, interestingly
  enough, it is found the disappearance of criticality as
  $\epsilon_{AA}\to 0$}. In our extension we can introduce an
additional parameter $s=\epsilon_{BB}/\epsilon_{AA}\red{>0}$. One
immediately verifies that the law
of mass action of Eq. (\ref{xaxb1})-(\ref{xaxb2}) admits now 4
solutions $(x_A,x_B)$ from which one has to determine the physical one
such that $x_A,x_B\in[0,1]$ and $\lim_{\rho\to 0}x_A=\lim_{\rho\to
  0}x_B=1$. Clearly in the limit $r\to 0$ the problem is similar to
the one of Bianchi {\sl et al.} \cite{Bianchi2006} (compare
Eqs. (\ref{xca})-(\ref{xcb}) and Eq. (\ref{xis})) and in the limit
$s\to 0$ we fall back to Russo {\sl et al.}
\cite{Russo2011a,Russo2011b} case. We are interested in the
non-trivial case: $\Delta_{AA}\neq \Delta_{BB}$ or
$M_A\neq M_B$. We will choose for $M_A,M_B,K_{AA}^0$ and
$K_{AB}^0$ the same values of the Russo's case of Section
\ref{sec:different-sites-binodal}. Moreover we will 
choose $K_{BB}^0=K_{AA}^0$. Again one has $\lim_{\rho\to 0}d\beta
p_{bond}^W/d\rho=0$.
For $s$ small we are still able to see the re-entrant liquid scenario
contrary to the predictions of Ref. \cite{Tavares2010}. In other words
we are able to observe a re-entrant liquid branch even in the presence
of X-junctions in the fluid, as long as the energy cost for their
formation, $\epsilon_{X-junction}=\epsilon_{AA}(1-s)$, is positive and
big
enough. This is shown in Fig. \ref{fig:rs1}. The figure also shows how
an ``R'' shaped spinodal is possible in these cases with a majority of
Y-junctions in correspondence of the coexistence region at high
temperature, a majority of X-junctions in correspondence of the
coexistence region at low temperature, and a majority of chains in
between in correspondence of the bottleneck in the ``R'', in agreement
with the study of Tavares {\sl et al.} \cite{Tavares2012b}. 
Moreover we find gas-liquid coexistence also for $r<1/3$ as long as
$s$ is large enough. This is shown in Fig. \ref{fig:rs2} from which it
is also apparent the existence of a gas-liquid coexistence with a
critical point at extremely low densities and temperatures, 
unpredicted by the study of Tavares {\sl et al.}
\cite{Tavares2012b}. As a matter of fact the critical temperature 
can be made small at will by a proper choice of the control
parameters $s$; the spinodal being essentially independent from
$r$.

%

\section{Break-down of the theory}
\label{sec:ltb}

Apart from the necessity to fulfill the steric incompatibility
conditions the Wertheim theory will break-down in the following cases:   
 
\subsection{Low temperature limit}

Both the Wertheim theory and the canonical Monte Carlo simulation
break-down at low temperatures. The Wertheim theory is a high 
temperature perturbation theory. The first order version that we have
been using until now clearly breaks-down at low temperature when from
the mass action law (\ref{x-wertheim}) follows that $x_\alpha\to 0$
which in turn produces an undefined bond free energy (\ref{bond-w}).
Also the usual Monte Carlo simulation will break-down at very low
temperatures. In fact, imagine we have to break a bond with a single
particle move. Then the total energy difference between the final
configuration and the initial one would be $\epsilon$ and we would
need around $1/e^{-\beta\epsilon}$ single particle moves. So at low
temperatures we would need a very long simulation in order to fully
explore configuration space. Depending from the computational
resources at one disposal the range of inaccessible temperatures,
before the solidification at zero temperature where the fluid chooses
spontaneously the minimum potential energy configuration, may vary. 
Even if it is possible that patchy fluids, with short-ranged and
tunable pair-interactions and with limited valence, will not
crystallize at zero temperature \cite{Smallenburg2013} remaining a
liquid in that limit.  

\subsection{Infinite number of attractive sites}

The Wertheim theory will not be applicable anymore to particles
decorated with too many attractive sites. In the limit of an infinite
number of sites uniformly distributed over the particle surface one
recovers the square-well fluid or the mean field solution of Section
\ref{sec:mf}. 

\section{The radial distribution function}
\label{sec:rdf}

Using the fact that the angular average of the functional derivative
of the free energy per particle respect to the angle dependent
pair-potential is equal to $\rho/2V$ times the radial
distribution function of colloid centers, we can write 
\red{
\bq
g(r)&=&g_0(r)+\frac{2V}{\rho}
\Big\langle\frac{\delta a_{bond}^W}{\delta\phi(1,2)}\Big\rangle\\
&=&g_0(r)+\frac{2}{\rho}\frac{1}{4\pi r^2}\sum_{\gamma\in\Gamma}
\left(\frac{1}{x_\gamma}-\frac{1}{2}\right)\Big\langle
\frac{\delta x_\gamma}{\delta\left[\sum_{\alpha,\beta\in\Gamma}
\beta\psi_{\alpha\beta}(r_{\alpha\beta})\right]}\Big\rangle, 
\eq
}
where we denote with $\langle\ldots\rangle$ the orientational average,
and in the second equality we used Eq. (\ref{bond-w}) and
Eq. (\ref{wertheim-potential}). 

To make some progress we use the following property
\red{
\bq 
\Big\langle\frac{\delta\langle
  f_{\alpha\beta}\rangle}{\delta\beta\psi_{\alpha\beta}}\Big\rangle=
-m_{\alpha\beta}(r)e^{\beta\epsilon_{\alpha\beta}}=
-m_{\alpha\beta}(r)-\langle f_{\alpha\beta}\rangle
\eq
}
\red{where in the last equality we used Eqs. (\ref{average-f}) and
  (\ref{m(r)})}. From Eq. (\ref{Delta-w}) follows  
\bq
\delta\Delta_{\alpha\beta}/\delta\langle 
f_{\alpha\beta}(12)\rangle = 4\pi
r_{12}^2g_0(r_{12})I_{\alpha\beta}(r_{12}),
\eq
where $I_{\alpha\beta}(r)$ is equal to one on the support of $\langle
f_{\alpha\beta}\rangle$ and zero otherwise. Next we observe that  
\red{
\bq \nonumber
\Big\langle\frac{\delta x_\gamma}{\delta\left[\sum_{\alpha,\beta\in\Gamma}
\beta\psi_{\alpha\beta}\right]}\Big\rangle&=&
\Big\langle\frac{1}{M^2}\sum_{\alpha,\beta\in\Gamma}\frac{\delta x_\gamma}
{\delta\beta\psi_{\alpha\beta}}\Big\rangle\\ \label{approx} 
&=&-4\pi r^2g_0(r)\frac{1}{M^2}\sum_{\alpha,\beta\in\Gamma}m_{\alpha\beta}(r)
e^{\beta\epsilon_{\alpha\beta}}\frac{\partial x_\gamma} 
{\partial\Delta_{\alpha\beta}},
\eq
}
where $M$ is the total number of sites per particle and in the last
equality we used the chain rule. So we obtain 
\bq \nonumber
&&g(r)=g_0(r)\Bigg[ 1+\\ \label{g1}
&&\left.\frac{1}{M^2\rho}\sum_{\alpha,\beta,\gamma\in\Gamma}
\left(1-\frac{2}{x_\gamma}\right)\frac{\partial
x_\gamma}{\partial\Delta_{\alpha\beta}}m_{\alpha\beta}(r)
e^{\beta\epsilon_{\alpha\beta}}\right],
\eq
where the terms $\frac{\partial
  x_\gamma}{\partial\Delta_{\alpha\beta}}$ can be determined from the
law of mass action, Eq. (\ref{x-wertheim}). In particular, using the
symmetry $\Delta_{\alpha\beta}=\Delta_{\beta\alpha}$, it follows
\bq \label{mag}
\frac{1}{\rho}\sum_{\gamma\in\Gamma}
\left(1-\frac{2}{x_\gamma}\right)
\frac{\partial x_\gamma}{\partial\Delta_{\alpha\beta}}
=x_\alpha x_\beta.
\eq
From Eq. (\ref{g1}) we can extract the contact
value for the radial distribution function  
\bq \nonumber
&&g(\sigma^+)=g_0(\sigma^+)\times\\
&&\left[1+\frac{1}{M^2\rho}\sum_{\alpha,\beta,\gamma\in\Gamma}
\left(1-\frac{2}{x_\gamma}\right)\frac{\partial
x_\gamma}{\partial\Delta_{\alpha\beta}}m_{\alpha\beta}(\sigma)
e^{\beta\epsilon_{\alpha\beta}}\right],
\eq
where $m_{\alpha\beta}(\sigma)$ is the product of the two solid angle
fractions for the $\alpha\beta$ bond when two particles are located at
relative center-to-center distance $\sigma$. For example for the Kern
and Frenkel pair-potential \cite{Kern03} we would have
$m_{\alpha\beta}=\chi_\alpha\chi_\beta$ with $\chi_\text{patch}$ the
patch surface coverage. In the Bianchi {\sl et al.} case
\cite{Sciortino2007} of Section \ref{sec:identical-sites} we have
instead $m_{\alpha\alpha}(\sigma)=(d/\sigma)^3/3$, from
Eq. (\ref{m(r)}). For $g_0(\sigma^+)$ we can use the analytic solution
to the Percus-Yevick approximation for the hard-sphere fluid
\cite{Hansen-McDonald-3}, namely  
\bq
g_0(\sigma^+)=(1+\eta/2)/(1-\eta)^2.
\eq
Next we observe that, since $\rho g(r)4\pi r^2dr$ gives the number of
particles in the spherical shell $[r,r+dr]$ around a particle fixed on
the origin, the coordination number can be estimated as follows  
\bq \nonumber
&&C_n=\rho\int_\sigma^{\sigma+d} 4\pi r^2g_0(r)\times\\
&&\left[1+
\frac{1}{M^2\rho}\sum_{\alpha,\beta,\gamma\in\Gamma}
\left(1-\frac{2}{x_\gamma}\right)\frac{\partial
x_\gamma}{\partial\Delta_{\alpha\beta}}m_{\alpha\beta}(r)
e^{\beta\epsilon_{\alpha\beta}}\right]\,dr,
\eq
where $d=\min\{d_{\alpha\beta}\}$. The mean number of bonds per
particle (the valence),
$v_T$ $=$ $\sum_{\alpha\in\Gamma}(1-x_\alpha)$, can be also 
estimated from the structure as follows
\bq \nonumber
v_S&=&C_n-\lim_{T\to\infty}C_n\\ \label{vS}
&=&\frac{1}{M^2}\sum_{\alpha,\beta,\gamma\in\Gamma}\left(1-\frac{2}{x_\gamma}\right)
\frac{\partial x_\gamma}{\partial\Delta_{\alpha\beta}}
\Delta_{\alpha\beta}.
\eq
Then using Eq. (\ref{mag}) we immediately find 
\bq
v_S=\frac{\rho}{M^2}\sum_{\alpha,\beta\in\Gamma}x_\alpha
x_\beta\Delta_{\alpha\beta}=\frac{1}{M^2}
\sum_{\alpha\in\Gamma}(1-x_\alpha),
\eq
where the last equality follows from the law of mass action,
Eq. (\ref{x-wertheim}). \red{The sought for consistency between the
  valence calculated from the thermodynamics and the valence
  calculated from the structure only holds in the single site
  per particle case, $M=1$}. 

For example, for $M$ identical sites we find $v_T=M(1-x)$ and,
choosing Kern-Frenkel patches for which $d$ represents the width of
the attractive square well of each patch and $\chi$ the patch surface 
coverage, from Eq. (\ref{mag}) follows  
\bq
C_n&=&\rho\int_\sigma^{\sigma+d} 4\pi r^2g_0(r)\left[1+
x^2\chi^2e^{\beta\epsilon}\right]\, dr.
\eq 

\section{The structure factor}
\label{sec:sk}

We then determined the structure factor $S(k)=1+\rho \hat{h}(k)$ with
$h(r)=g(r)-1$ the total correlation function and the hat denotes the
Fourier transform. 

\subsection{Identical sites}
\label{sec:sis}

For the case of Bianchi {\sl et al.} of Section
\ref{sec:identical-sites} we find
\bq \nonumber
S(k)&=&1+4\pi\rho\int_0^\infty
\left\{g_0(r)\left[1+x^2m(r)e^{\beta\epsilon}\right]-1\right\}\times\\
\label{sk}
&&\frac{\sin(kr)}{k}r\,dr,
\eq
where $x$ is given by Eq. (\ref{xis}) and $m(r)$ is given by
Eq. (\ref{m(r)}). 
Choosing for $g_0(r)=\Theta(r-\sigma)$ the one obtained from the zero
density limit of the hard-sphere fluid, we find the ``triangular''
approximation result of Eq. (\ref{sk1}) of Appendix \ref{app:1}. From
this result follows immediately 
\bq \nonumber
&&\lim_{k\to 0}S(k)=1+\\ \nonumber
&&20\eta\bigg[\left(e^{\beta\epsilon}-8M\eta(e^{\beta\epsilon}-1)\right)
(15d^4+4d^5)-\\ \nonumber
&&4\left(5+\sqrt{5}\sqrt{5+4d^4M\eta(e^{\beta\epsilon}-1)(15+4d)}\right)
\bigg]/\\ \label{s0s}
&&\left(5+\sqrt{5}\sqrt{5+4d^4M\eta(e^{\beta\epsilon}-1)(15+4d)}\right)^2. 
\eq
Moreover we find
\bq
\lim_{T\to 0}S(0)&=&1-8\eta+\frac{1}{M},\\
\lim_{T\to\infty}S(0)&=&1-8\eta+\left(3d^4+\frac{4}{5}d^5\right)\eta,
\eq
whereas for the structure factor of the reference system we have
$S_0(0)=1-8\eta$. 

\begin{figure}[htbp]
\begin{center}
\includegraphics[width=8cm]{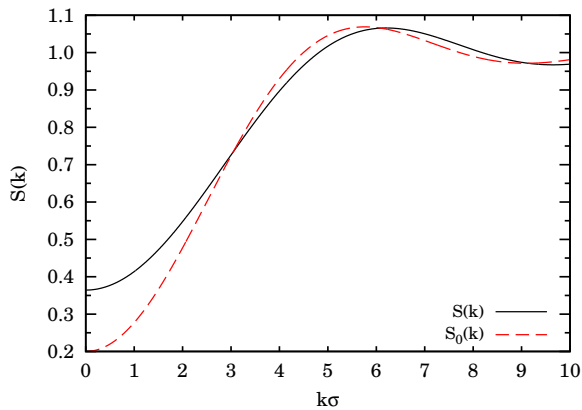}
\end{center}  
\caption{(color online) Structure factor for $M=4$ and
  $T^*=0.1,\eta=0.1$ in the Bianchi {\sl et al.} case using for the
  radial distribution function of the reference system, $g_0$,
  the zero density limit of the hard-sphere fluid. Also shown, for
  comparison, is the structure factor of the reference system,
  $S_0(k)=1+24\eta(k\cos(k)-\sin(k))/k^3$.}
\label{fig:sf}
\end{figure}

In Fig. \ref{fig:sf} we show the structure factor of Eq. (\ref{sk1})
for $M=4$ and $T^*=0.1, \eta=0.1$. 

\begin{figure}[htbp]
\begin{center}
\includegraphics[width=8cm]{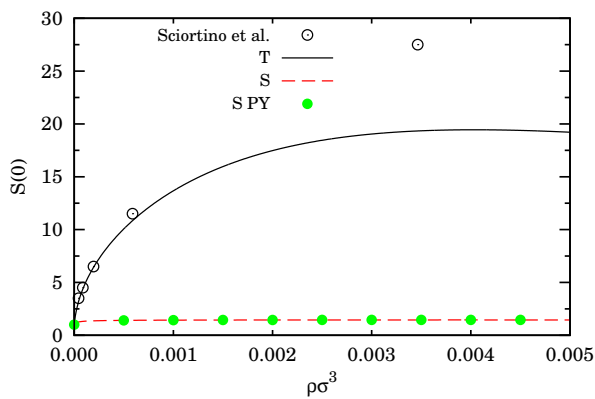}
\end{center}  
\caption{(color online) Structure factor at zero wave-number as a
  function of density for $M=2$ and $T^*=0.055$ in the Sciortino {\sl
    et al.} simulations of Ref. \cite{Sciortino2007}, from the
  thermodynamic route (T) of the isothermal compressibility of
  Eq. (\ref{s0t}), from the structure route (S) of Eq. (\ref{s0s}),
  and from the zero wave-number limit of Eq. (\ref{sk}) taking as a
  reference system the Percus-Yevick analytic solution for
  hard-spheres (S PY).}  
\label{fig:s0}
\end{figure}

A comparison with the simulation results of Sciortino {\sl et al.}
\cite{Sciortino2007} (see their Fig. 13) at $M=2$ and $T^*=0.055$
shows that approximation (\ref{s0s}) breaks-down at high
densities. This is shown in Fig. \ref{fig:s0} where the data of
Sciortino {\sl et al.} simulations are compared with the isothermal
compressibility sum rule,
\bq \label{s0t}
S(0)=\left[\frac{\partial}{\partial\rho}\left(
\rho^2\frac{\partial\beta a}{\partial\rho}\right)\right]^{-1},
\eq 
and the relationship between the activity $\Lambda^{-3}e^{\beta\mu}$
and the density is obtained through Eq. (\ref{bm}). We think that the
fact that the structure as determined by the Eq. (\ref{sk}) does not
satisfy the isothermal compressibility sum rule of Eq. (\ref{s0t}) is
a thermodynamical inconsistency not universally recognized for the
Wertheim theory. In order to find accurate results for the structure
one needs to solve the Wertheim Ornstein-Zernike equation with an
appropriate closure \cite{Chang1995}.

%

\section{Conclusions}
\label{sec:conclusions}

We have critically analyzed some recent applications of the 
Wertheim perturbation  theory 
to classes of associating fluids of with non standard phase diagrams
and increasing
complexity which can be today
engineered in the laboratory \cite{Yi2013}.  In particular, we have 
illustrated the strong structural stability of the theory, which allows to get a first
correct qualitative understanding of the resulting phase diagrams, even at the simplest level
where all correlations of the reference system are neglected. 

For fluids of hard-spheres with $M$ identical bonding sites Bianchi
{\sl et al.} \cite{Bianchi2006,Sciortino2007,Bianchi2008} discovered the
``empty liquid'' scenario as $M$ approaches two, {\sl i.e.} in the presence
of ``chains'' only. The phenomenology when there are sites of two
different kinds is more rich \cite{Russo2011a,Russo2011b} and one can
have ``junctions'', responsible for a re-entrance of 
the liquid branch of the binodal, and ``rings''
\cite{Tavares2012,Rovigatti2013}, responsible for a
re-entrance also in the gas branch and the appearance of a second
lower critical point.

In our detailed analysis of these results, we show that all the 
important conclusions on the qualitative behaviour of the phase
diagrams can be derived uniquely from theoretical analytical considerations 
without the need of inputs from simulation results. For example, for
the case of rings forming fluids we used as the partition function of
an isolated ring the Treloar analytic expression for a freely jointed
chain, unlike Rovigatti {\sl et al.} \cite{Tavares2012,Rovigatti2013}
who use a fit of the MC data. This approximation
makes immediately available a useful tool of analysis of complex phase diagrams 
even in absence of more accurate
but heavy numerical results. 

Also in the case of the more demanding condition of the presence
of   X-junctions we find that, when the
energy gain for an X-junction formation, $s$, is low enough, we still
observe a re-entrant liquid branch for $r<1/2$ in the fluid,
eventually with an ``R'' shaped spinodal in agreement with the study
of Tavares {\sl et al.} \cite{Tavares2012b}. When $s$ is sufficiently
large we observe gas-liquid coexistence also at $r<1/3$ in agreement
with the predictions of Ref. \cite{Tavares2010}. In these latter cases a
gas-liquid coexistence with a critical point at an extremely low
density and temperature, unpredicted by the work of Tavares {\sl
    et al.} \cite{Tavares2012b}, can be observed. 

Moreover, 
we have discussed in detail the 
consistency between structural and thermodynamic description within Wertheim 
perturbation theory and in particular the valence as obtained from the
thermodynamics and from the structure. We can conclude that while the
overall structural information underlying the first order perturbative
level is not accurate, the theory provides \red{a consistency condition on the}
estimate of bonded particles, \red{which is satisfied only in the one-site case}.
An analytical expression for
the radial distribution function and the structure factor has also been
proposed.

\appendix
\section{The structure factor in the ``triangular'' approximation} 
\label{app:1}

Choosing $g_0(r)=\Theta(r-\sigma)$ in Eq. (\ref{sk}) with $m(r)$
defined as in Eq. (\ref{m(r)}), we find
\bq \nonumber
&&S(k)=\\ \nonumber
&&1+80\eta\bigg[(15k^3-90d^4k^3M\eta-24d^5k^3M\eta)\cos(k)+\\ \nonumber
&&(90d^4k^3M\eta+24d^5k^3M\eta+10d^3k^3)e^{\beta\epsilon}\cos(k)+\\ \nonumber
&&3\sqrt{5}k^3\sqrt{5+4d^4M\eta(e^{\beta\epsilon}-1)(15+4d)}\cos(k)+\\ \nonumber
&&(-15k^2+90d^4k^2M\eta+24d^5k^2M\eta)\sin(k)+\\ \nonumber
&&(-90d^4k^2M\eta-24d^5k^2M\eta)e^{\beta\epsilon}\sin(k)+\\ \nonumber
&&(15d^2k^2+30)e^{\beta\epsilon}\sin(k)+\\ \nonumber
&&-3\sqrt{5}k^2\sqrt{5+4d^4M\eta(e^{\beta\epsilon}-1)(15+4d)}\sin(k)+\\ \nonumber
&&30(dk\cos(k(1+d))-\sin(k(1+d)))e^{\beta\epsilon}\bigg]/\\ \label{sk1}
&&\left[k^5\left(5+\sqrt{5}\sqrt{5+4d^4M\eta(e^{\beta\epsilon}-1)(15+4d)}
\right)^2\right].
\eq
\red{From this expression one immediately sees that the high temperature
limit, $\beta\to 0$, of the structure factor is independent from the
number of sites, $M$.} 


\subsection{Acknowledgements}
We are grateful to Jos\'e Maria Cantista de Castro Tavares for
correspondence and helpful comments. G.P. acknowledges financial support by
PRIN-COFIN 2010-2011
(contract 2010LKE4CC).
\bibliographystyle{tMPH}
\bibliography{MPassociation}
\end{document}